\documentclass[a4paper,11pt]{article}

\usepackage{graphicx}
\usepackage{graphics}
\usepackage{jcappub} 

\usepackage{jcappub} 

\usepackage{makecell}
\usepackage{xcolor}
\usepackage{caption}
\usepackage{siunitx}
\usepackage{subfigure}
\usepackage{subfloat}
\usepackage{float}
\usepackage{multirow}
\usepackage{rotating}
\usepackage{pdflscape}
\usepackage{tabulary}  
\usepackage{booktabs}
\usepackage[T1]{fontenc} 
\usepackage{gensymb}
\title{\boldmath Performance and systematic uncertainties of CTA-North in conditions of reduced atmospheric transmission}

\author[a,b]{Mario Pecimotika,\note{Corresponding author.}}
\author[b]{Dijana Dominis Prester,}
\author[c,1]{Dario Hrupec,}
\author[b]{Saša Mićanović,}
\author[b]{Lovro Pavletić,}
\author[d]{and Julian Sitarek}

\affiliation[a]{Ruđer Bošković Institute, \\ Bijenička cesta 54, Zagreb, Croatia}
\affiliation[b]{University of Rijeka, Faculty of Physics, \\ Radmile Matejčić 2, Rijeka, Croatia}
\affiliation[c]{J. J. Strossmayer University of Osijek, Department of Physics, \\ Trg Ljudevita Gaja 6, Osijek, Croatia}
\affiliation[d]{University of Łódź, Department of Astrophysics, \\ Pomorska 149, Łódź, Poland}

\emailAdd{mario.pecimotika@irb.hr}
\emailAdd{dijana@phy.uniri.hr}
\emailAdd{dario.hrupec@fizika.unios.hr}
\emailAdd{sasa.micanovic@phy.uniri.hr}
\emailAdd{lovro.pavletic@phy.uniri.hr}
\emailAdd{julian.sitarek@uni.lodz.pl}

\abstract{The Cherenkov Telescope Array (CTA) is the next-generation stereoscopic system of Imaging Atmospheric Cherenkov Telescopes (IACTs). In IACTs, the atmosphere is used as a calorimeter to measure the energy of extensive air showers induced by cosmic gamma rays, which brings along a series of constraints on the precision to which energy can be reconstructed. The presence of clouds during observations can severely affect Cherenkov light yield, contributing to the systematic uncertainty in energy scale calibration. To minimize these systematic uncertainties, a calibration of telescopes is of great importance. For this purpose, the influence of cloud transmission and altitude on CTA-N performance degradation was investigated using detailed Monte Carlo simulations for the case where no action is taken to correct for the effects of clouds. Variations of instrument response functions in the presence of clouds are presented. In the presence of clouds with low transmission ($\leq 80\%$) the energy resolution is aggravated by 30\% at energies below 1 TeV, and by 10\% at higher energies. For higher transmissions, the energy resolution is worse by less than 10\% in the whole energy range. The angular resolution varies up to 10\% depending both on the transmission and altitude of the cloud. The sensitivity of the array is most severely reduced at lower energies, even by 60\% at 40 GeV, depending on the clouds' properties. A simple semi-analytical model of sensitivity degradation has been introduced to summarize the influence of clouds on sensitivity and provide useful scaling relations.}


\begin{document}
\maketitle
\flushbottom
\section{Introduction}

Very-high-energy (VHE, energy
 $\gtrsim 100$ GeV) gamma rays have proven crucial for probing the most extreme cosmic phenomena. To date more than 200 VHE sources have been observed, such as active galactic nuclei, pulsar wind nebulae, supernovae remnants, gamma-ray bursts, and binary systems\footnote{\href{http://tevcat.uchicago.edu/}{http://tevcat.uchicago.edu/}}.
Direct detection of cosmic gamma rays relies on space-borne experiments, a technique inefficient in the VHE regime due to low flux and limited collection area. Although the atmosphere is nontransparent to gamma rays, they can be detected indirectly from the ground. The ground-based experiments \cite{Sitarek:2022cpk} make use of the extensive air showers, a cascade of secondary particles produced in the interaction of gamma rays with molecules in the atmosphere. The cascade starts with the pair production mechanism. As the primary gamma ray interacts with the electric field of the atmospheric nucleus, it is transformed most often into an electron-positron pair which may, in turn, produce secondary gamma rays through bremsstrahlung. The process is repeated iteratively, rapidly increasing the number of secondary particles, until the energy at which the ionization losses overcome radiative energy losses ($\approx$ 85 MeV in the air) \cite{Matthews:2005}. 

The Whipple collaboration, which led to the discovery of the TeV emission from the Crab Nebula in 1989 \cite{Weekes:1989tc} using the Imaging Air Cherenkov Technique (IACT), can be considered a pioneer in the field of ground-based VHE gamma-ray astronomy. IACT is a widely used method for indirect observation of secondary particles produced in air showers, based on imaging short flashes of Cherenkov light produced by ultra-relativistic charged particles. The Cherenkov telescopes have a field of view (FoV) of a few degrees, a gamma-ray angular resolution of the order of 0.1$\degree$, and fast cameras composed of $\approx 10^3$ photomultipliers. The gamma-ray image in the camera has an approximately two-dimensional elliptical angular distribution of the captured Cherenkov light, with the major and minor axes following the longitudinal and lateral development of the shower, characterizing the nature of the primary particle. The reconstruction of energy and arrival direction can be improved by stereoscopic observations. The three major, currently operating stereoscopic IACT systems are the High Energy Stereoscopic System (H.E.S.S.) \cite{Ashton:2020zic}, the Major Atmospheric Gamma Imaging Cherenkov telescopes (MAGIC) \cite{Aleksic:2014poa}, and the Very Energetic Radiation Imaging Telescope Array System (VERITAS) \cite{Holder:2006}.

The Cherenkov Telescope Array (CTA) is the future ground-based observatory for gamma-ray astronomy at very-high energies which will consist of the northern (La Palma, Spain) and southern (Paranal, Chile) arrays. CTA will cover a broad energy range from 20 GeV to 300 TeV by combining three types of telescopes, providing a sensitivity of $\approx$ 0.1\% of the Crab Nebula flux in 50 hours of observations\cite{Knodlseder:2020onx, CTAObservatory:2022mvt}. The Large-Sized Telescopes (LSTs) with a parabolic reflective surface of 23 m in diameter and FoV of 4.3$\degree$ are designed to be dominant in an energy range between 20 GeV and 150 GeV. The Medium-Sized Telescopes (MSTs), a 12 m Davies-Cotton optical design with FoV of 7$\degree$, are optimized in the core energy range from 150 GeV to 5 TeV. An array of Small-Sized Telescopes (SSTs) with a 4.3 m Schwarzchild-Couder optical system and a large FoV of 8.8$\degree$ will provide sensitivity at energies up to 300 TeV. After the initial construction phase, the Northern CTA Observatory (CTAO-North, CTA-N) will cover the area of $\approx 0.25$ km$^2$ gathering in operation 4 LSTs and 9 MSTs\footnote{This is the so-called \textit{Alpha Configuration}, currently the official CTA Observatory configuration. However, this paper is based on the \textit{Omega Configuration} (section \ref{sec:mc}),  which refers to the full-scope design possibly implemented depending on the available funds.}. The Southern CTA Observatory (CTAO-South, CTA-S) will span over 3 km$^2$ combining 14 MST and 37 SST telescopes. A unique operational design and a wide energy range will allow CTA to achieve extraordinary sensitivity, becoming a leading observatory in the branch of ground-based VHE gamma-ray astronomy.

The atmosphere is a target medium for VHE gamma rays because Cherenkov light is produced and transmitted through the atmosphere and affects the Cherenkov yield in two ways. First and foremost, the amount of Cherenkov light produced in the air shower depends on the refractive index of the air. At small core distances, the lateral density of Cherenkov light is strongly dependent on the atmospheric density profile and shows variations of 15\%-20\% between different profiles (absorption included) for mid-latitudes \cite{Bernl_hr_2000, Bernlohr:2014upa}, although more recent studies \cite{Munar:2019} suggest that the effect is an order of magnitude smaller at CTA sites. In addition, the Cherenkov light may be absorbed or scattered in poor atmospheric conditions, reducing the total number of Cherenkov photons recorded by the telescope camera. While molecular (Rayleigh) scattering is well-known and almost independent of geographical location, ozone absorption and scattering by clouds and aerosols are time-variable and depend on location and pointing position \cite{Bernl_hr_2000,Dubovik:2006}.

The duty cycle of IACTs depends considerably depends on the state and quality of the atmosphere, as well as on the lunar phases. While it is relatively easy to manage the lunar influence \cite{Ahnen:2017, Cortina:2015}, it is of utmost importance to understand the performance of Cherenkov telescopes in conditions of reduced atmospheric transmission. The influence of clouds on IACT was studied in \cite{Sobczynska:2014dba,Sobczynska:2013aya,DOROTA2020102450,2013ICRC...33.2909G,Devin:2019}. The presence of clouds affects the images of gamma-induced and hadron-induced showers in a similar way. For example, in the presence of low and dense clouds, the angular distance between the camera center and the image center of gravity, the so-called \textit{dist} parameter, is shifted towards higher values.  The images in the camera captured in the presence of clouds tend to be broader. The images may be distorted as well, reducing the efficiency of gamma-hadron separation. A reduced number of Cherenkov photons decreases the trigger rate of the telescope, inducing an energy bias and decreasing effective collection area \cite{Nolan2010xe,ConsortiumCarloVigoritofortheCTA:2017yfy}. However, the effects are small and negligible when the shower maximum lies below the clouds.

In the past, the quality of the atmosphere was judged by the rate of cosmic rays. Over time, more precise and sophisticated methods and strategies for atmospheric calibration of Cherenkov telescopes have been developed. Various instruments such as LIDAR, UAV-based systems, ceilometer, pyrometer, and the All-Sky camera are used to characterize the state of the atmosphere and calibrate the telescopes \cite{Daniel:2015zda,Doro:2015nba,Brown:2018,Brown:2022,CTAConsortium:2017tgk, Will:2017hnj}. Several methods have been developed to enable the analysis of cloud-affected data. Most methods rely on measuring atmospheric parameters with different instruments and correcting the data. For example, a correction method presented in \cite{DOROTA2020102450} is based on the fraction of Cherenkov light produced above the cloud compared to the total Cherenkov light produced in the shower and the energy bias calculated for a clear atmosphere. The results are consistent with simulations in the energy range between 2 TeV and 30 TeV with systematic uncertainties of less than 20\% for a transmission of 0.6 or higher. In another method \cite{Fruck:2014mja, Fruck:2015mba}, the corrected energy is obtained as a reconstructed energy scaled by the inverse of the average optical depth.

The paper presents a comprehensive study on the influence of reduced atmospheric transmission on the performance of CTA-N and its corresponding subarrays. The study is based on atmospheric modeling and Monte Carlo simulations (MC) of air showers and telescope responses.

\section{Monte Carlo simulations}
\label{sec:mc}

\subsection{Air shower and detector simulations} 
Telescope calibration and optimization are to a great extent based on the MC simulations. The first step is common to all astroparticle physics experiments and consists of air shower simulations based on the COsmic Ray SImulations for KAscade (CORSIKA) code \cite{heck1998corsika, corsika}, a standard MC tool in the IACT community. Produced secondary particles are tracked through the atmosphere and subjected to further interactions, decays, and emissions of Cherenkov light. While electromagnetic interactions are treated with EGS4 code \cite{osti_6137659}, a choice between various hadronic interaction models is given. CORSIKA simulations are time-consuming and require lots of disk space. The IACT/ATMO package is a mandatory extension in the simulation of the Cherenkov light emission and its transport through the atmosphere. The package is used to configure the geometry of the telescope array, atmospheric transmission profiles, atmospheric refraction, observation level, etc. To reduce the size of the output file each shower is reused at different random impact points from the shower core, and only photons intersecting the fictional spheres around any of the telescopes are stored.

\begin{table}[t]
\centering
\begin{tabular}{@{}llll@{}}
\toprule
                    & \textbf{gamma (on-axis)} & \textbf{proton} & \textbf{electron} \\ \midrule
ERANGE (TeV)        & 0.003 - 330                               & 0.003 - 600                      & 0.004 - 600                        \\
NSHOW               & 2$\cdot 10^8$                             & 4$\cdot 10^9$                    & 2$\cdot 10^8$                      \\
VIEWCONE (deg)      & 0                                         & 10                               & 10                                 \\
CSCAT         & 10                      & 20             & 20                \\
maximum impact (m)         & 1400                       & 1900           & 1900             \\
zenith (deg)  & 20                                        & 20                               & 20                                 \\
azimuth (deg) & 180                                       & 180                              & 180                                \\ \bottomrule
\end{tabular}
\caption{Basic CORSIKA parameters: ERANGE - the energy range; NSHOW - the number of simulated air showers; VIEWCONE - the direction of a primary particle from a cone around a pointing zenith and azimuth angle; CSCAT - the number of reuses of the same air shower. \label{tab:corsika_details}}
\end{table}

The CORSIKA version 7.6400 and hadronic models UrQMD \cite{Bass_1998} and QGSJET-II-04 \cite{Ostapchenko_2006} were used in the study. Air showers induced by cosmic gamma rays and the background induced by protons and electrons arriving from the geographical South were simulated. Gamma-ray showers were simulated at the center of the camera while the background was simulated arriving from a cone with a half-opening angle of 10$\degree$ around the same zenith and azimuth angle (details are presented in table \ref{tab:corsika_details}). Cherenkov light has been simulated in the wavelength range from 240 nm to 700 nm. Simulations have been performed for CTA-N observatory (2147 m a.s.l., $\ang{28;45;43.7904}$ N, $\ang{17;53;31.218}$ W).

In the second part of MC production, another tool called \texttt{sim\_telarray} is used \cite{corsika}. In this stage, the atmospheric impact on the propagation of Cherenkov photons as well as the response of the Cherenkov detector is simulated. The software simulates optical ray-tracing, electronic response, additional NSB and electronic noise, triggers, and in the final step computes pixels response. Each detected photo-electron results in an identical analog signal of a different amplitude. 

\begin{figure}[t] 
    \centering
    \includegraphics[width=0.65\textwidth]{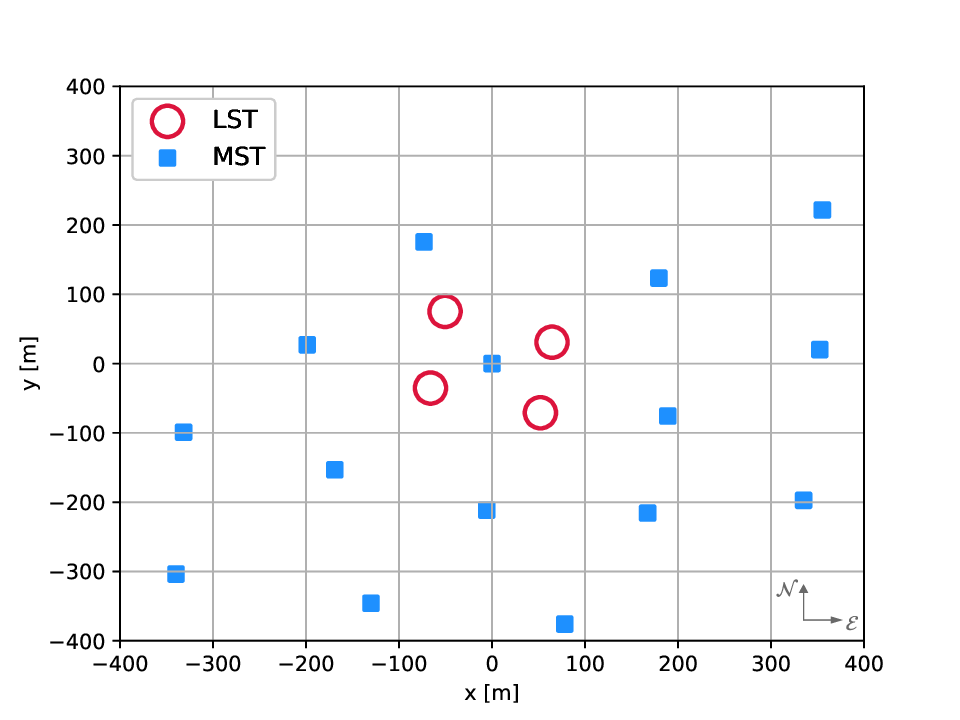}
    \caption{Simulated array configuration for the Northern Cherenkov Telescope Array consists of 4 Large-Sized Telescopes and 15 Medium-Sized Telescopes.}
    \label{fig:layout}
\end{figure}

The CORSIKA output is directly piped to  \texttt{sim\_telarray} simulations using the so-called \texttt{multipipe\_corsika} option. The \texttt{multipipe\_corsika} option enables simultaneous simulation of different configurations of telescope layouts or transmission profiles. The option is especially efficient for this study given a large number of different atmospheric conditions. In
general, \texttt{sim\_telarray} requires only a fraction of the total CPU time needed for simulations. The final output is a
full digitized pulse shape for every pixel. Once the data is generated, MC simulations are finalized. Simulated telescope array (4 LSTs and 15 MSTs, the so-called \textit{Omega Configuration}) is shown in figure \ref{fig:layout}.

\subsection{Atmospheric modelling}
As explained in the previous section, \texttt{sim\_telarray} stage of MC simulations is responsible for the simulation of the atmospheric effects on Cherenkov light propagation. A user has a choice to use the default \texttt{sim\_telarray} transmission profile or provide a custom one that contains an atmospheric optical depths as a function of wavelength and altitude in a tabular form.  The modeling of the atmospheric conditions has been done using the MODerate resolution atmospheric TRANsmission (MODTRAN) code version 5.2.2 \citep{Berk1987MODTRAN:LOWTRAN,Berk2005MODTRANUpdate,stotts2019atmospheric}. MODTRAN treats the atmosphere as a set of homogeneous layers and, based on the input settings, calculates atmospheric optical properties by solving the radiative transfer equation.  Atmospheric transmission and emission calculations are based on predefined (e.g. tropical atmosphere, mid-latitudes atmosphere, U.S. Standard Atmosphere, etc.) or user-defined models of the atmosphere, aerosol profiles, water clouds, and ice cloud contents.

The atmospheric transmission in the presence of 1 km thick altostratus clouds in the wavelength range from 200 nm to 1000 nm (in steps of 1 nm) was calculated. The predefined atmospheric model U.S. Standard Atmosphere \cite{1992P&SS...40..553N} was used as a basis, with the desert aerosol extinction model of default visibility of 75.748 km for the boundary layer (from 0 to 2 km altitude). The default spring-summer seasonal dependence of the U.S Standard Atmosphere is chosen for tropospheric aerosols for layers from 2 to 10 km, and the background stratospheric aerosol profile for layers above 10 km. The model gives a total aerosol optical depth of $\approx$ 0.16 at 550 nm, which is in agreement with previous studies of aerosol profiles over the North Atlantic \cite{Barreto2022,Freville2020,Gallo2023,Kinne2006} and a relative difference of only 1.43\% compared to the official atmospheric model for CTA-N used in \texttt{sim\_telarray}. 

A recent study by \cite{Fruck2022CharacterizingLIDAR} found that most of the cloud cover over La Palma consists of single-layer clouds. The cloud bases during spring and winter were typically found between 8 and 10 km above a.s.l., while summer clouds were on average found at 8 km a.s.l. In general, the distribution of clouds over La Palma extends between 4 and 14 km a.s.l., although occasionally clouds occur even in the stratosphere (up to 22 km a.s.l.), with the lowest obtained total vertical transmission of $\approx 40\%$. It is important to note that the observations are not performed in the presence of lower clouds. Therefore, the data for lower clouds may be underestimated and should be interpreted with caution. Accordingly, the transmission profiles were simulated for the atmosphere with the cloud base at six different altitudes above ground level (a.g.l., from 3 km to 13 km in steps of 2 km) and six different transmissions $T$ (0.95, 0.90, 0.80, 0.75, 0.60, 0.50). The cloudless atmosphere ($T = 1$), used as a benchmark in the assessment of the influence of clouds on telescope performance,  has been modeled as well.

\begin{figure}[t]
    \centering
    \includegraphics[width=0.65\textwidth]{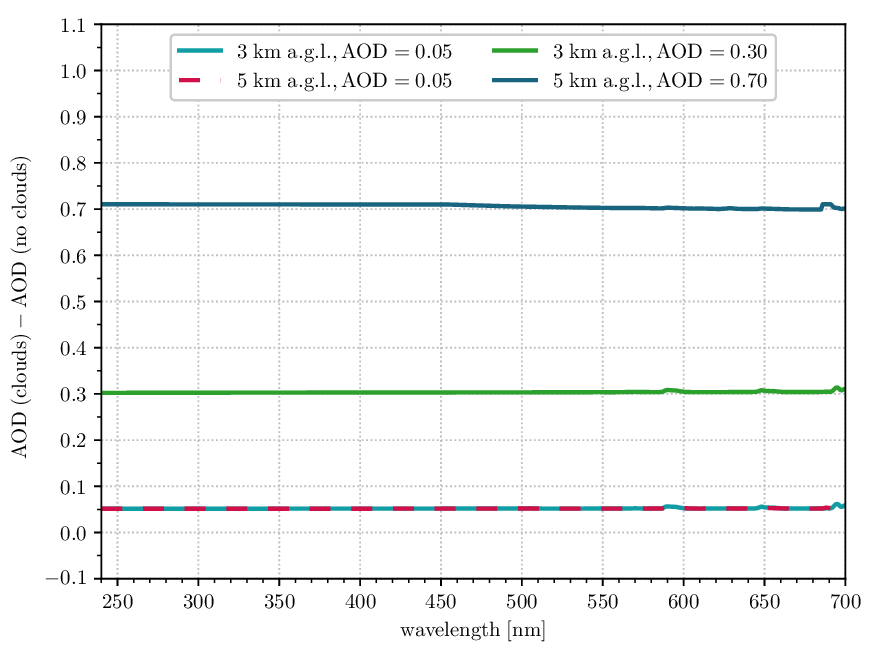}
    \caption{Difference between total aerosol optical depth in the presence of clouds and total aerosol optical depth in cloudless conditions from 9 km a.g.l to the ground for selected cases. Note the wavelength-independent AOD difference.}
    \label{fig:aod_diff}
\end{figure}

The optical properties of clouds have been studied extensively in the past, focusing on the wavelength dependence of optical depth \cite{Fu1996, Hu1993, Min2004, Serrano2015}. According to calculations based on Mie scattering theory, optical depth variations are typically $\lesssim$ 2\% over the wavelength range from 300 to 1000 nm, and are considered to be nearly independent of wavelength. Therefore, the simulated clouds are nearly grey, while the internal structure of the cloud is neglected, assuming uniform extinction throughout the cloud. Figure \ref{fig:aod_diff} shows a difference between the total aerosol optical depth (AOD) in the presence of clouds and total AOD in cloudless conditions from 9 km a.g.l. to the ground for selected cases. The AOD difference is shown as a function of wavelength in the wavelength range of the Cherenkov light simulated in CORSIKA. For example, the mean value for clouds at an altitude of 3 km a.g.l. with AOD = 0.05, corresponding to $T$ = 0.95, is 0.0519 $\pm$ 0.0012. On the other hand, the mean value for cloud at an altitude of 5 km a.g.l. with AOD = 0.70 ($T$ = 0.50) is 0.7359 $\pm$ 0.0044. The results show that the simulated cloud optical depths are indeed independent of wavelength. Multiple scattering was not considered as it is not important for IACTs due to the limited and small FoV \cite{Bernl_hr_2000}.

\section{Analysis method}

\subsection{Signal extraction and calibration}
The simulated data were analyzed using the MAGIC Analysis and Reconstruction Software (MARS) \cite{Zanin:2013,Sitarek:2018,Moralejo:2009}.  In cases where the energy reconstruction of data taken in the presence of clouds is based on the MCs for a clear atmosphere, a bias in reconstructed energy may be introduced \cite{Devin:2019,DOROTA2020102450,Nolan2010xe}. A self-consistent analysis was performed in this study given that the same kind of MCs was used as a training sample and a test sample to avoid any additional biases.  

The \texttt{sim\_telarray} output files contain the waveform of the signal for each event and each pixel, and they need to be processed with the Convert Hessio Into Mars inPut (\texttt{chimp}) to obtain an appropriate input format for MARS analysis. \texttt{chimp} is a MARS-based program that integrates the waveforms using a pixel-wise peak-search algorithm. The pixel signals are reduced to two numbers: charge and arrival time. The \texttt{chimp} performs two-pass signal extraction for each triggered telescope using the sliding window algorithm for the first pass, and the fixed window extractor for the second pass \cite{Aleksic:2016}, see also \cite{Albert:2008, Sitarek:2018}. 

In the first pass, for each pixel, the algorithm is searching the maximum sum of three consecutive time slices using a fixed readout window through the whole range of the readout slices. After the maximum sum of three consecutive time slices has been found, two additional samples, one per each side of the search window, are summed to obtain the total pixel charge.  The signal is calculated by subtracting the mean pedestal value (the baseline of the waveform previously calculated by \texttt{sim\_telarray}) from the region with the highest readout sum. The mean arrival time in readout samples is determined as the average of the readout slices' time, weighted over the readout counts.

In the second pass, a time gradient along the major axis is calculated on the preliminary cleaned image as described in the section \ref{sec:image_cl_para}. In this step, the signal is searched in non-significant pixels (all pixels except the so-called core pixels in the main island of the preliminary image) in a reduced, fixed time window (5 samples again), obtained from the time fit. If the predicted time for a pixel is outside the readout window, the first or last five samples are integrated, recovering some dimmer signals surrounding the main island. The image obtained after the second-pass waveform integration is cleaned again, using final cleaning cuts (section \ref{sec:image_cl_para}, Table \ref{tab:cleaning}) and, eventually, parameterized.

\begin{table}[t]
\centering
\begin{tabular}{@{}lcccc@{}}
\toprule
\textbf{}               & \multicolumn{2}{c}{\textbf{preliminary cleaning}} & \multicolumn{2}{c}{\textbf{final cleaning}} \\ \midrule
                        & $Q_c$ (phe)             & $Q_b$ (phe)             & $Q_c$ (phe)          & $Q_b$ (phe)          \\
\multicolumn{1}{c}{LST} & 6                       & 3                       & 4                    & 2                    \\
\multicolumn{1}{c}{MST} & 8                       & 4                       & 4                    & 2                    \\ \bottomrule
\end{tabular}
\caption{\label{tab:cleaning} Boundary pixels ($Q_b$) and core pixels ($Q_c$) image cleaning thresholds expressed in photoelectrons (phe). Preliminary cleaning is performed in the second pass of the signal extraction to recover some weak signals. After the final cleaning, the surviving pixels are used to parameterize the image.}
\end{table}

\subsection{Image cleaning and parametrization}
\label{sec:image_cl_para}


The image cleaning process is performed to suppress signals in the camera which are induced by the night sky background (NSB) or signals arising from the electronic noise. The standard cleaning algorithm is the two-level absolute image cleaning, a simple and robust method that discriminates pixels according to their charge, classifying them as core pixels or boundary pixels \cite{Aliu:2009, Sitarek:2018}. The core pixel is a pixel with a charge above a certain threshold $Q_c$ with at least one neighboring pixel which satisfies the threshold value. The boundary pixels are pixels with the charge above the threshold charge $Q_b$ with at least one neighboring core pixel (cleaning thresholds are presented in table \ref{tab:cleaning}). All other pixels are excluded from further steps of the analysis. This method does not achieve the lowest possible energy threshold as the arrival time information is not used, but sufficiently high thresholds provide good discrimination between the NSB signals and signals induced by air showers. The cleaned image is parametrized to the so-called Hillas parameters \cite{Hillas:1985}.

\subsection{Reconstruction of arrival direction}
Instead of using all triggered telescopes to characterize an event, a set of quality cuts is applied to not use poorly reconstructed showers. To be used for the stereoscopic reconstruction of events, the image recorded by the telescope must satisfy different requirements. For example, the \textit{size}, which is defined as the total charge contained within the shower image, must exceed 50 phe. The distance between image center of gravity and the camera center must be less than 82\% of the camera radius. Furthermore, reconstructed \textit{impact parameter} $<$ 200 m, the number of events in the corresponding look-up table bin $\geq$ 10, and $0.1 < \textit{width over length} < 0.6$ are required as well. The default cut values set in the software workflow were used in this study. 

The primary gamma-ray direction is calculated using the so-called look-up tables. The look-up tables are two-dimensional histograms filled with the square of mean \textit{miss} in bins of \textit{size} and \textit{width over length}. In this way, for particular values of \textit{size}, \textit{length}, and \textit{width}, it can be estimated how well the image axis is aligned with the true gamma-ray direction in the camera. In the stereoscopic reconstruction of the events with at least three valid images, the values stored in the look-up tables are used to set relative weights for different telescopes. The direction is calculated as the point in the camera that minimizes the weighted sum of squared distances between the source and the shower axis. Shower maximum height and core position are calculated in a similar manner.

\subsection{Reconstruction of energy}
The energy reconstruction is performed using the Random Forest (RF) algorithm \cite{Albert:2008b}. The RF is trained on the gamma train subsample and built per telescope type. For every event, and for each valid image, energy is calculated under the assumption the image in the camera is from the gamma-initiated shower. The RF consists of 50 trees. To build a node, three randomly selected parameters are considered of which the one providing the minimal dispersion of the logarithm of true energy $\log E_\mathrm{true}$ is chosen. The splitting is done when the final node content is 5 events. The RF output is estimated or reconstructed energy $E_\mathrm{est}$ and the root mean square ($\mathrm{RMS}$) resulting from all the trees. The final reconstructed energy is calculated as a weighted average of the $E_\mathrm{est}$ obtained by $n$ telescopes which have survived image cleaning, parametrization, and quality cuts, where $\mathrm{RMS}^{-2}$ is used as a weight:

\begin{equation}
    E_\mathrm{est}=\frac{\sum_{i=1}^{n} E_{\mathrm{est},i}\cdot \mathrm{RMS}_i^{-2}}{\sum_{i=1}^{n}\mathrm{RMS}_i^{-2}}.
\end{equation}

\subsection{Gamma-hadron separation}
\label{sec:g/h}
During the observations, Cherenkov telescopes are not only triggered by Cherenkov photons produced in the electromagnetic cascades, but also by those produced in the hadron-initiated showers, giving an important yield to the background. Hadronic showers are not the only contribution to the background, e.g. the telescopes may be triggered by the background light fluctuations \cite{Gaug:2013, Compagnino:2022ugq} or single muons (also originating from hadronic showers) \cite{Gamez:2020,Aab_2015,Nieto:2021}. However, the signal fluctuations due to background light have a very low trigger rate (few Hz) and they are well managed in the low-level analysis (image cleaning). The muon trigger rate is up to five times smaller than the hadron rate, which makes hadronic showers the most important component of the background. For this matter, a procedure of discrimination between gamma-initiated and hadron-initiated events is needed.

Background rejection is performed by using the RF algorithm \cite{Albert:2008b,Bernloehr:2012}. The result of this procedure is a single number called hadronness, $h \in \left[0, 1\right]$. The hadronness indicates how likely is that the recorded event is of hadronic origin - the closer the hadronness value is to 1, the event is more likely of hadronic origin. The tree growing begins with the complete sample contained in a single node. In the node, three highly discriminating parameters are randomly chosen and considered for a cut. Out of the considered parameters, the one which better splits the sample is used. In this way, the initial sample is divided into two parts, the so-called branches. The procedure is repeated iteratively until a node contains only one type of event.  The global hadronness $h$ of the event is calculated as a weighted average of the hadronness $h_i$ calculated for different telescopes, where $size_i^{0.54}$ is used as a weight (the value of the exponent is obtained empirically):

\begin{equation}
    h=\frac{\sum_{i=1}^{n} h_{i}\cdot size_i^{0.54}}{\sum_{i=1}^{n}size_i^{0.54}}.
\end{equation}

\section{CTA-North performance in the presence of clouds}

\subsection{Differential sensitivity}
The differential sensitivity is the minimum intrinsic flux that can be detected with a statistical significance of 5$\sigma$ (using eq. (17) from \cite{1983ApJ...272..317L}) from a source with a power-law spectrum similar to the Crab Nebula in a defined observational time (50 hours in this paper). The differential sensitivity is calculated in non-overlapping logarithmic energy bins (five per decade), with a minimal number of 10 gamma-ray counts per energy bin, and a background-to-signal ratio of 0.05.

\begin{figure}[t] 
    \centering
    \centerline{
    \subfigure[4 LSTs] 
        {
            \includegraphics[width=0.32\textwidth]{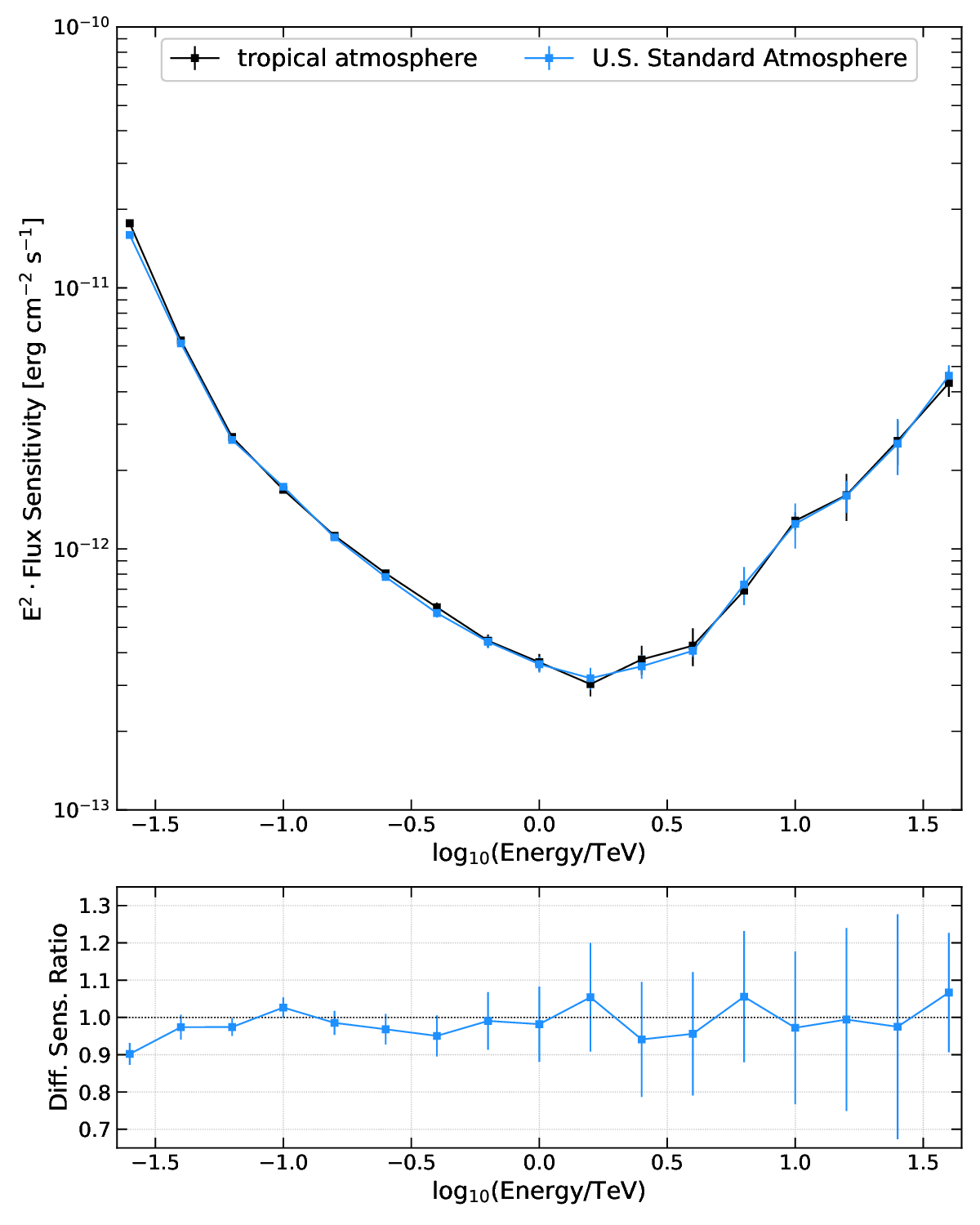}
        } 
    \subfigure[15 MSTs] 
        {
            \includegraphics[width=0.32\textwidth]{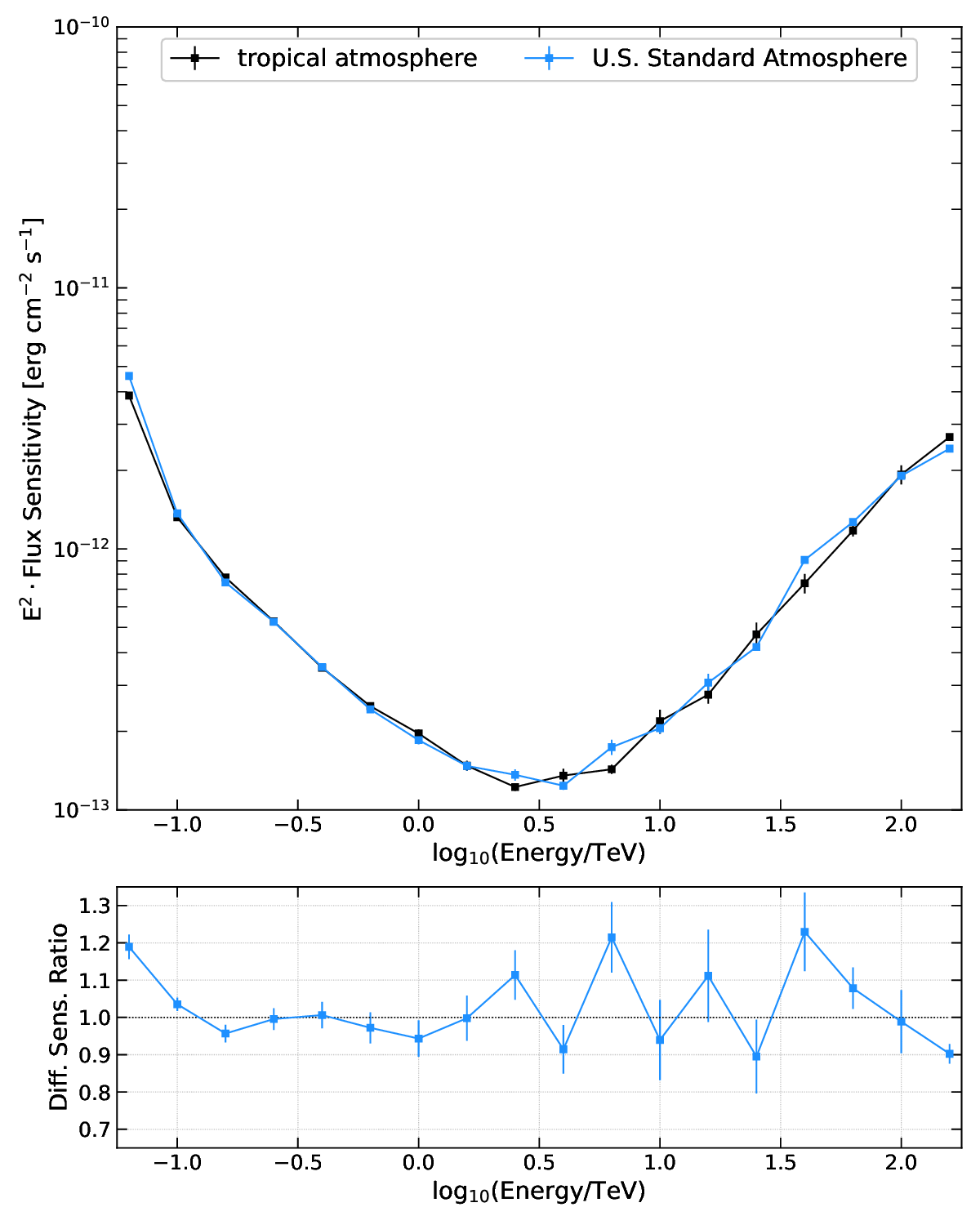}
        } 
    \subfigure[CTA-N] 
        {
            \includegraphics[width=0.32\textwidth]{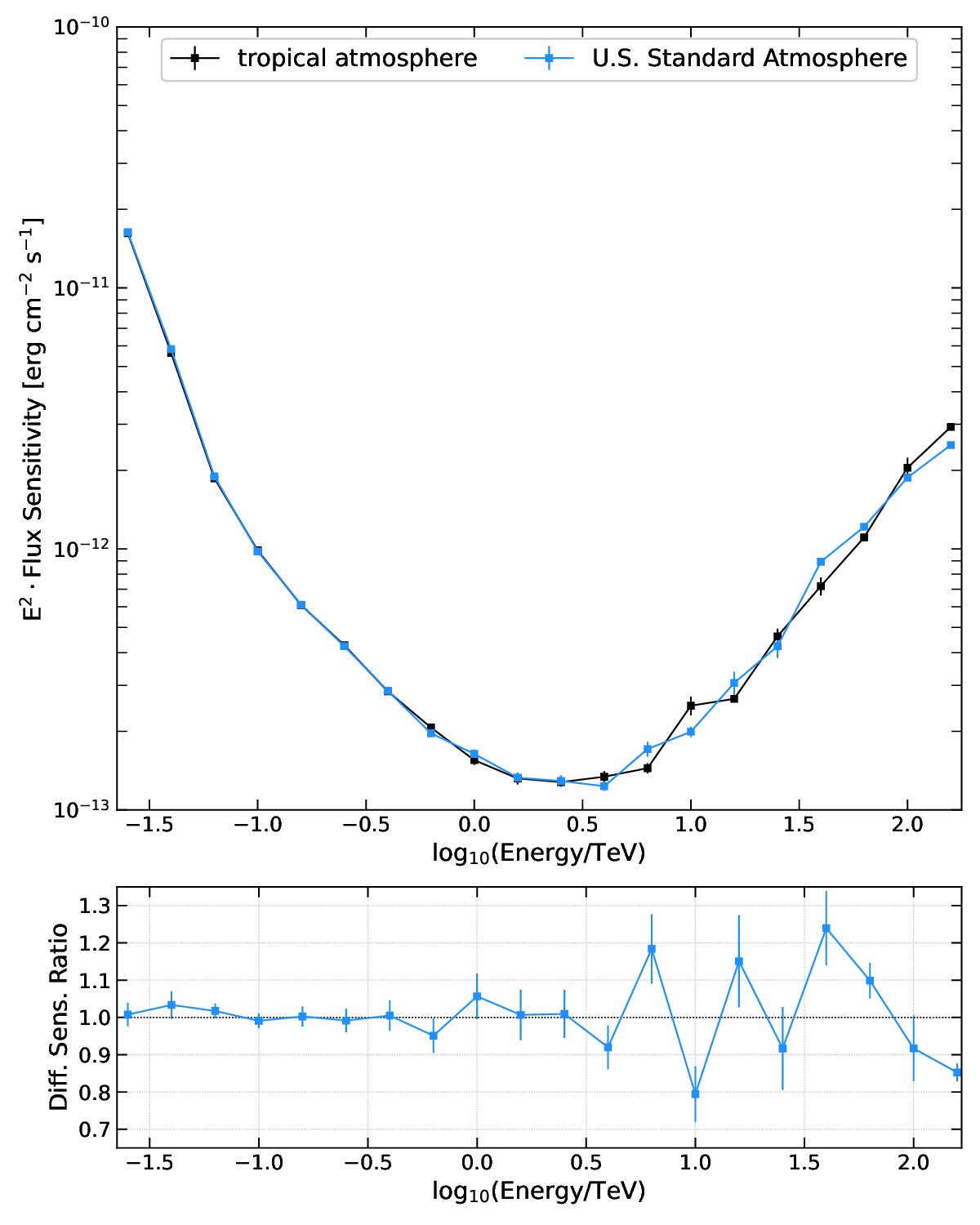}
        } 

}
    \caption{Differential sensitivity and differential sensitivity ratio as a function of energy for the subarray of 4 LSTs (left panel), subarray of 15 MSTs (middle panel) and full CTA-N array (right panel) - a comparison between the tropical atmosphere (the default \texttt{sim\_telarray} model) and the U.S. Standard Atmosphere (this study). Vertical error bars indicate the uncertainty in the sensitivity.}
    \label{fig:simtel}
\end{figure}

The standard atmospheric model used for \texttt{sim\_telarray} simulations in CTA Consortium is the tropical atmosphere. However, the U.S. Standard Atmosphere was used for this study given the well-known characteristics of its mathematical model. Deviations in the sensitivity in the case of 4 LSTs, 15 MSTs, and full CTA-N arising from different atmospheric models are presented in figure \ref{fig:simtel}. In the case of 4 LSTs, the relative differences in sensitivity do not exceed 10\%. In the full sensitivity range from 20 GeV to 150 GeV, the differences are $\lesssim$ 5\%. A similar trend is evident for the subarray of 15 MSTs. In the full sensitivity range for MSTs (150 GeV - 5 TeV), the ratio values are in general $\lesssim$ 10\%. Regarding the CTA-N layout, the differences are $\lesssim$ 5\% for energies below 3 TeV. At energies $\geq$ 3 TeV, there is a performance drop due to the poor sensitivity of 4 LSTs at higher energies.

\begin{figure}[t] 
    \centering
    \subfigure[3 km a.g.l.] 
            {
                \includegraphics[width=0.45\textwidth]{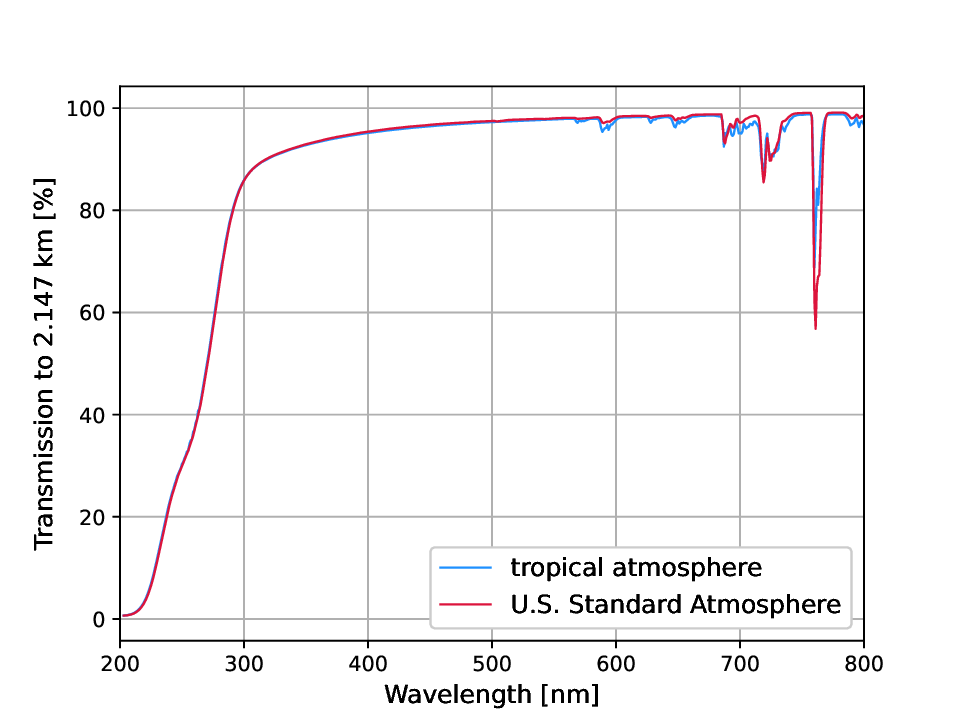}
            } 
    \subfigure[5 km a.g.l.] 
            {
                \includegraphics[width=0.45\textwidth]{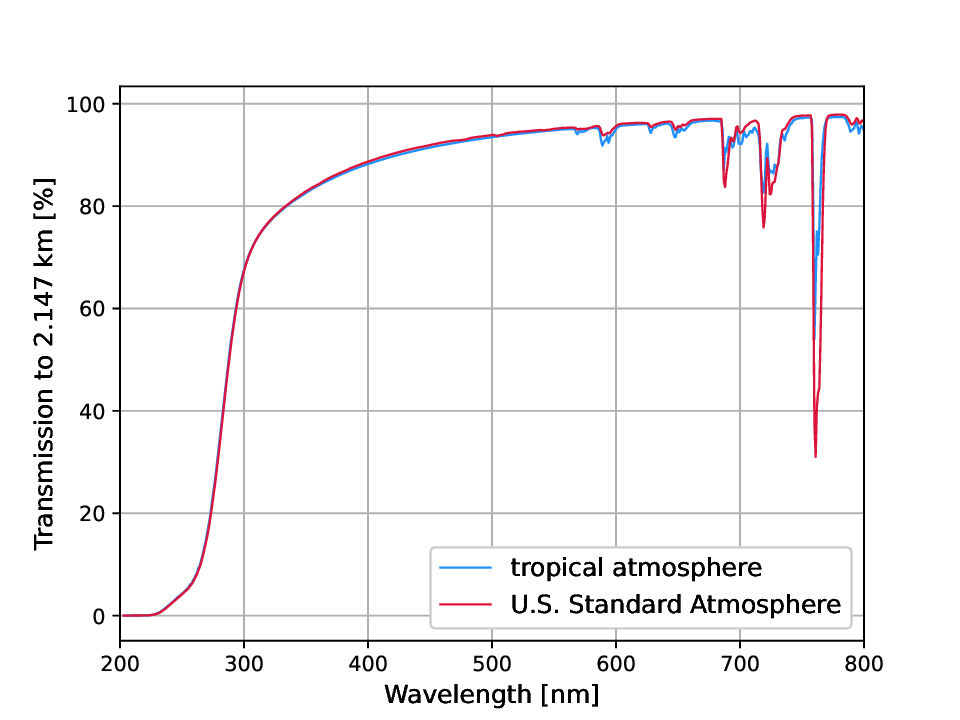}
            } 
     \subfigure[7 km a.g.l.] 
            {
                \includegraphics[width=0.45\textwidth]{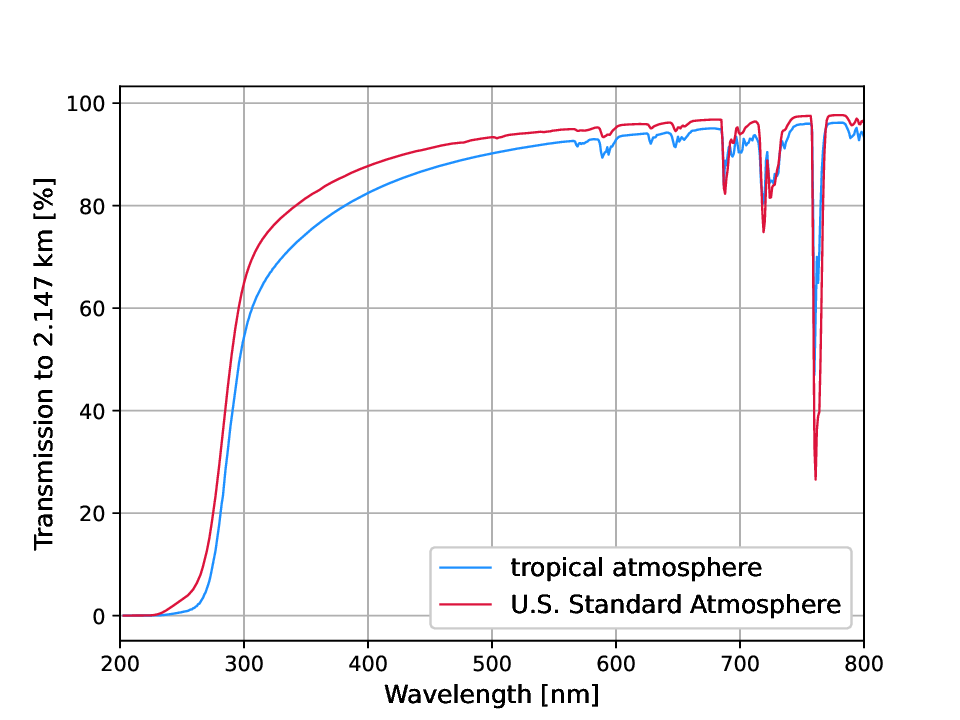}
            } 
    \subfigure[9 km a.g.l.] 
            {
                \includegraphics[width=0.45\textwidth]{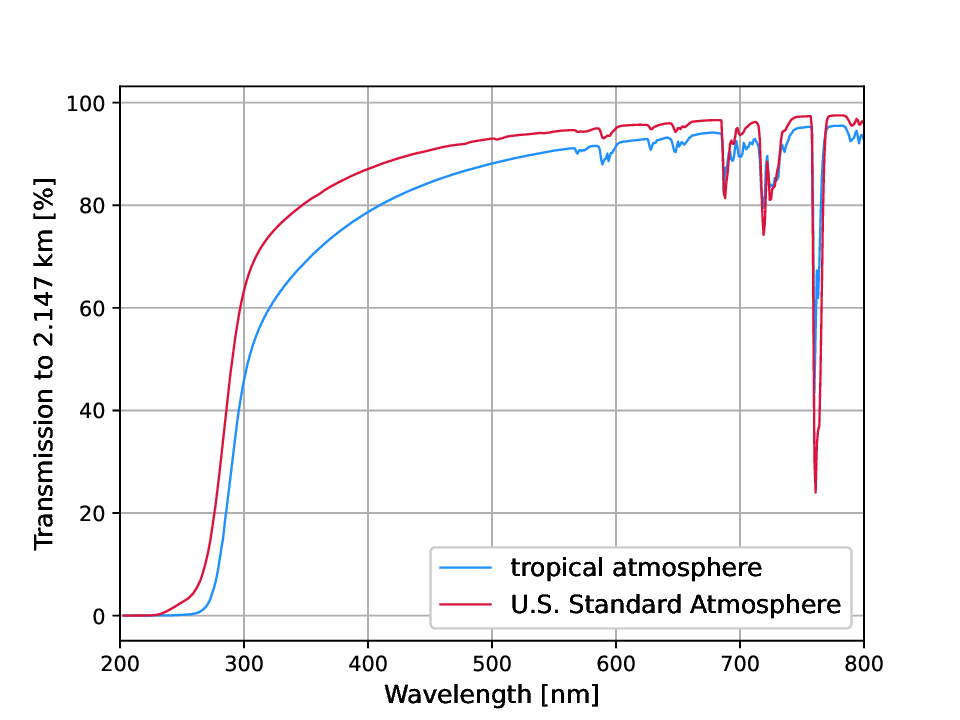}
            } 
    \caption{Comparison of atmospheric transmission at 3, 5, 7, and 9 km a.g.l. (the ground level is 2147 m above the sea level) for two different atmospheric models - the tropical atmosphere (blue line) and the U.S. Standard Atmosphere (red line). Both models are obtained by MODTRAN. The models differ most significantly in their temperature, water, and ozone content, while other constituents have similar density profiles. The differences in transmissions for altitudes below 7 km a.g.l., in the wavelength range relevant for Cherenkov telescopes, amount to a few percent.  At higher altitudes, the differences tend to be higher as there is intrinsically more light absorption.}
    \label{fig:atmotrans}
\end{figure}

Given the nature of MC and RF methods, these differences may stem from simulations and analysis rather than from the models themselves as the difference between the models (in the wavelength range significant for Cherenkov telescopes) are negligible for altitudes below 7 km a.g.l. (figure \ref{fig:atmotrans}). The differences between transmission profiles are greater at higher altitudes, as there is intrinsically more absorption. However, differences do not introduce any systematic errors as all data is processed in the same way. The deviations in the sensitivity may result from the differences in the effective areas\footnote{In general, the effective area is defined as the geometrical area around the telescope where a gamma-ray shower produces a trigger, folded with the gamma-ray efficiency of all the cuts applied in the analysis \cite{Aleksic:2016}. Detection efficiency is defined as a ratio of gamma-ray events that survived all cuts to the total number of simulated events. } and residual cosmic-ray background rates as well, as they are dependent on the quality cuts which are optimized to achieve the best sensitivity in each energy bin. The best sensitivity is achieved through large effective areas or low background, therefore, small fluctuations in the simulated data may result in large apparent fluctuations \cite{Bernloehr:2012}. 

Regarding the comparison of the differential sensitivity for a clear atmosphere and the atmosphere with clouds, the differences in the sensitivity increase with decreasing altitude and transmission of the cloud. Low and dense clouds ($\leq7$ km a.g.l., $T\leq0.75$) have the most prominent effect of the different sensitivity of the subarray of 4 LSTs (figure \ref{fig:subfigsLSTsdiff}). For example, in the case of clouds at 3 and 5 km a.g.l. and transmission $T = 0.50$, telescope sensitivity is worse by $\approx$ 40\% at 40 GeV, becoming more stable at higher energies ($\lesssim$ 20\%). Clouds with transmission $T = 0.95$ seem to have a negligible ($\lesssim$ 5\%) effect on the performance regardless of their altitude. Clouds of intermediate transmission ($T = 0.90$ and $T = 0.80$)  at energies above 100 GeV spoil sensitivity by $\lesssim$ 10\%.

In the case of 15 MSTs (figure \ref{fig:subfigsMSTsdiff}), the clouds at $\leq7$ km a.g.l. for which $T\leq0.75$ worsen sensitivity across the entire energy range, although by only 5\% at highest energies. For example, at the energy threshold, the sensitivity is worse by $\approx$ 80\% in the case of clouds at 3 km a.g.l. and $T = 0.50$, while the effect is smaller in the presence of more transparent clouds ($T = 0.75$),  reaching the relative difference of $\approx$ 25\%. In general, the influence of clouds with $T>0.75$ on the sensitivity of MSTs, as well as very high clouds ($\geq 11$ km a.g.l.) of lower transmissions, is small over the entire energy range. At energies $\gtrsim$ of a few hundred GeV, the sensitivity is worse by up to 30\% in the presence of clouds at $\leq9$ km a.g.l., or by 10\% in the case of higher clouds.

Clouds have a smaller effect on the sensitivity of CTA-N (figure \ref{fig:subfigsCTANdiff}) compared to the sensitivity of 4 LSTs and 15 MSTs separately. However, the influence of clouds with $T\leq 0.8 $ is still large at energies below 2.5 TeV. In the case of clouds at 3 km a.g.l. the sensitivity of CTA-N at an energy of 40 GeV is decreased by $\approx$ 60\% in the worst-case scenario. At energies $\gtrsim$ of a few hundred GeV, the reduction in sensitivity is stable, and throughout the energy range, the relative difference between the sensitivity in the presence of clouds and the sensitivity in a clear atmosphere does not exceed 20\%. Selected cases are also listed in the table \ref{tab:diffsens}.

\subsubsection{Semi-analytical model of sensitivity degradation}
\label{sec:semi}
To summarize the influence of clouds on the sensitivity, and to provide useful scaling relations, a simple semi-analytical model of the sensitivity degradation has been introduced. The expected height of the shower maximum a.g.l. is estimated as:
\begin{equation}
    H_\mathrm{max}=-H_\mathrm{obs} - H_0\cdot \ln \bigg(\ln \frac{E_\mathrm{est}}{80\,\mathrm{MeV}}\bigg)\cdot \frac{X_0}{X_\mathrm{all}},
\end{equation}
where $H_0=8$ km is the atmosphere exponential scale, $H_\mathrm{obs}=2.2$ km is the observation level of the telescopes, $X_0=38$ $\mathrm{g\,cm^{-2}}$ is the cascade unit, and $X_\mathrm{all}=10^3$ $\mathrm{g\,cm^{-2}}$ is the full thickness of the atmosphere.  The above equation assumes a low zenith angle of the observations.  

The fraction of the light attenuated by the cloud with transmission $T$ is evaluated by comparison of $H_\mathrm{max}$ with the altitude of the cloud, $H$:
\begin{equation}
    \eta=(1-T)\cdot\Sigma(H_\mathrm{max} - H, B),
\end{equation}
where $\Sigma(x, s)=0.5\cdot(\tanh(x\cdot s)+1)$ is a sigmoid function used to smooth the dependence, and $B$ is a free parameter of the model.

With the loss of Cherenkov light, the performance is rapidly degraded near the energy threshold.  Therefore, a threshold effect coefficient $\zeta$ is introduced:
\begin{equation}
    \zeta=\Sigma\bigg(\log\frac{E_\mathrm{est}(1-\eta)}{\mathrm{TeV}} - D, A\bigg).
\end{equation}
While parameter $D$ can be interpreted as the threshold of the system without the cloud presence, it is partially degenerated with the smoothening parameter $A$. 

Finally, the sensitivity in the presence of clouds $S(T,H)$ is given by:
\begin{equation}
    S(T, H) = \frac{S(T=1)}{\zeta \bigg(1-\bigg[C+F\cdot\log\frac{E_\mathrm{est}}{\mathrm{TeV}}\cdot\eta\bigg]\bigg)},
    \label{eq:dsr}
\end{equation}
where $S(T=1)$ is the sensitivity of the telescopes in a clear atmosphere (calculated from the MC simulations), while $C$ and $F$ are parameters of a linear function (of logarithmic energy) describing the proportionality of the fraction of lost Cherenkov light to a drop in performance. The parameters of the model are listed in table \ref{tab:model} for all three arrays considered.

\begin{figure*}[t]
        \centering
            \subfigure[3 km a.g.l.] 
            {
                \label{subfig:lab1}
                \includegraphics[width=.45\textwidth]{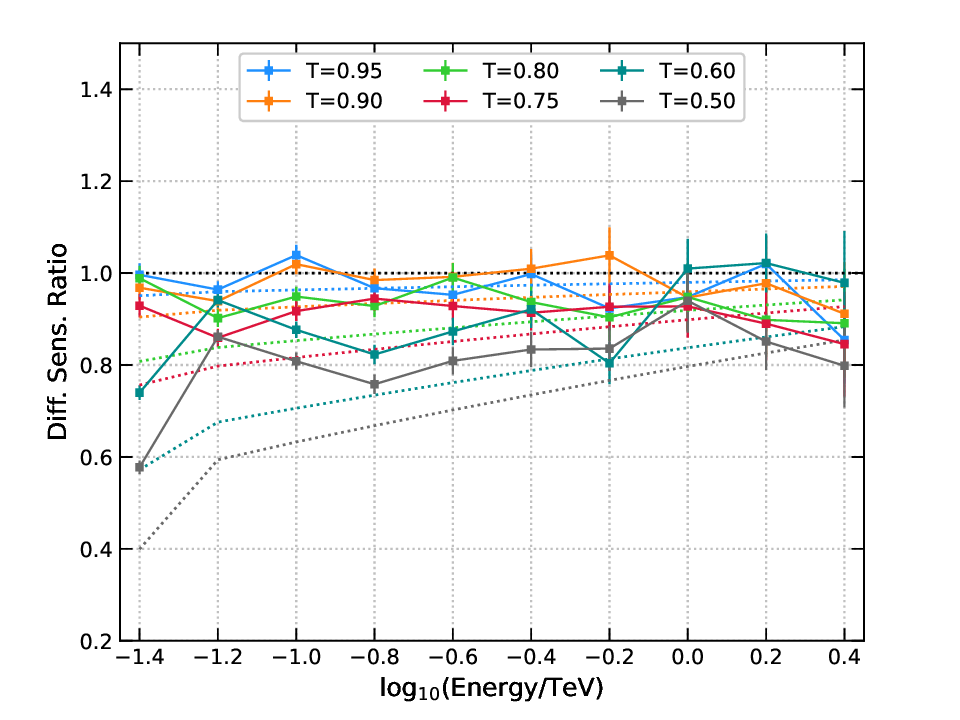} 
            } 
            \subfigure[5 km a.g.l.] 
            {
                \label{subfig:lab2}
                \includegraphics[width=.45\textwidth]{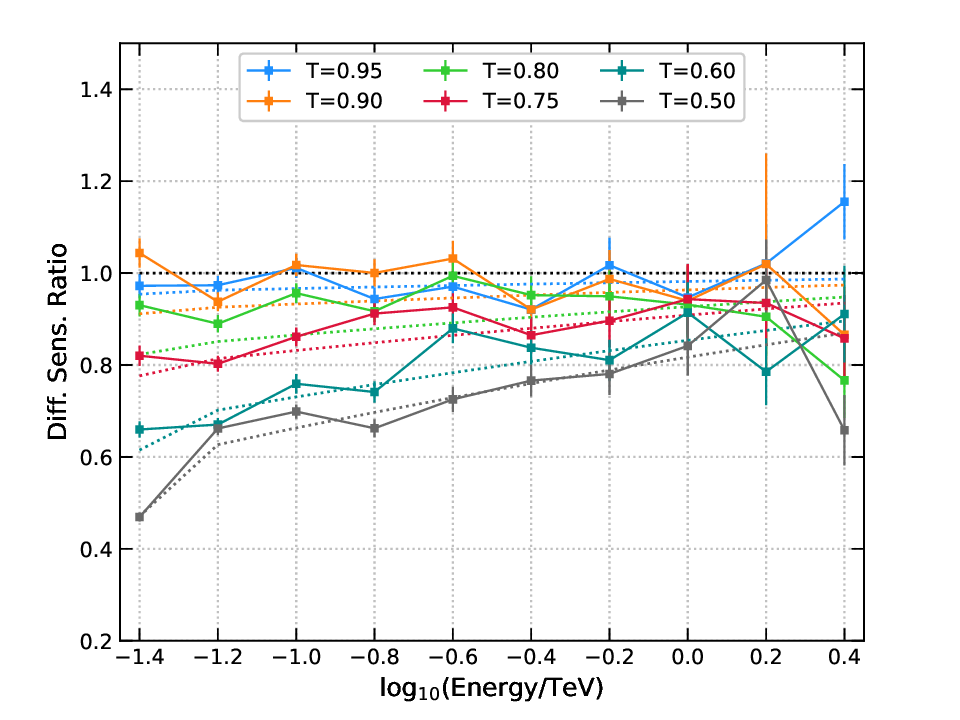} 
            }\\
            \subfigure[7 km a.g.l.] 
            {
                \label{subfig:lab3}
                \includegraphics[width=.45\textwidth]{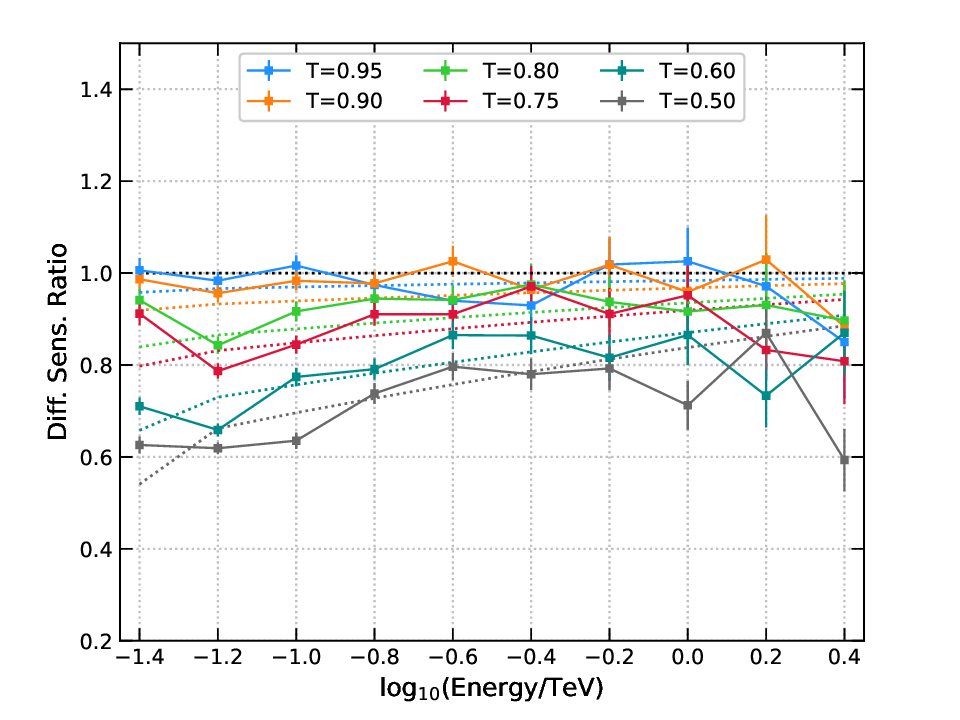}
            } 
            \subfigure[9 km a.g.l.] 
            {
                \label{subfig:lab4}
                \includegraphics[width=.45\textwidth]{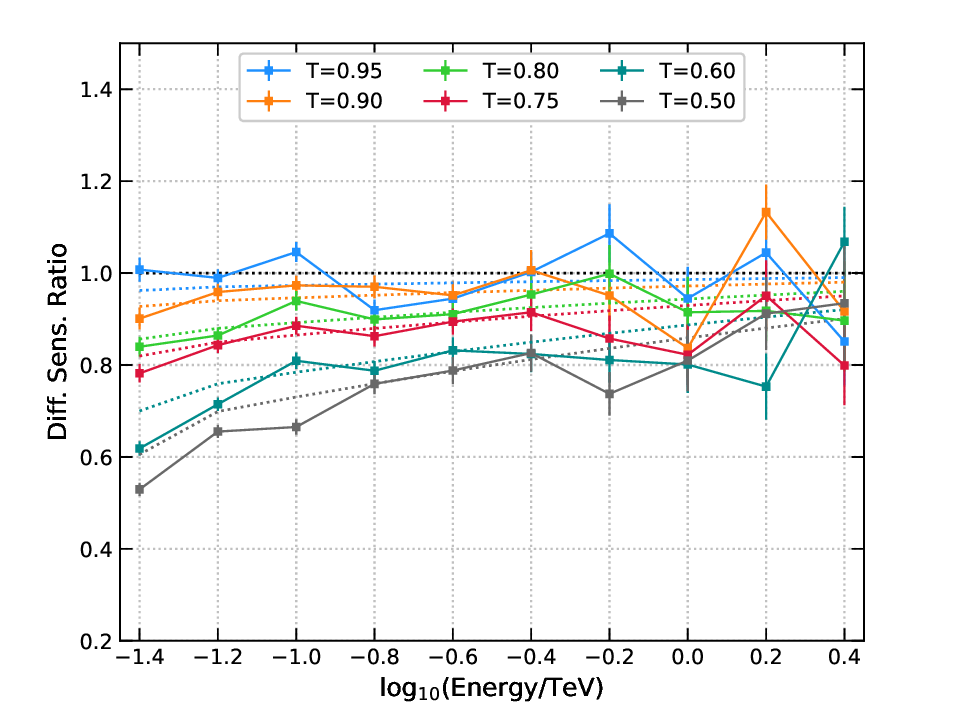} 
            }\\
            
            \subfigure[11 km a.g.l.] 
            {
                \label{subfig:lab5}
                \includegraphics[width=.45\textwidth]{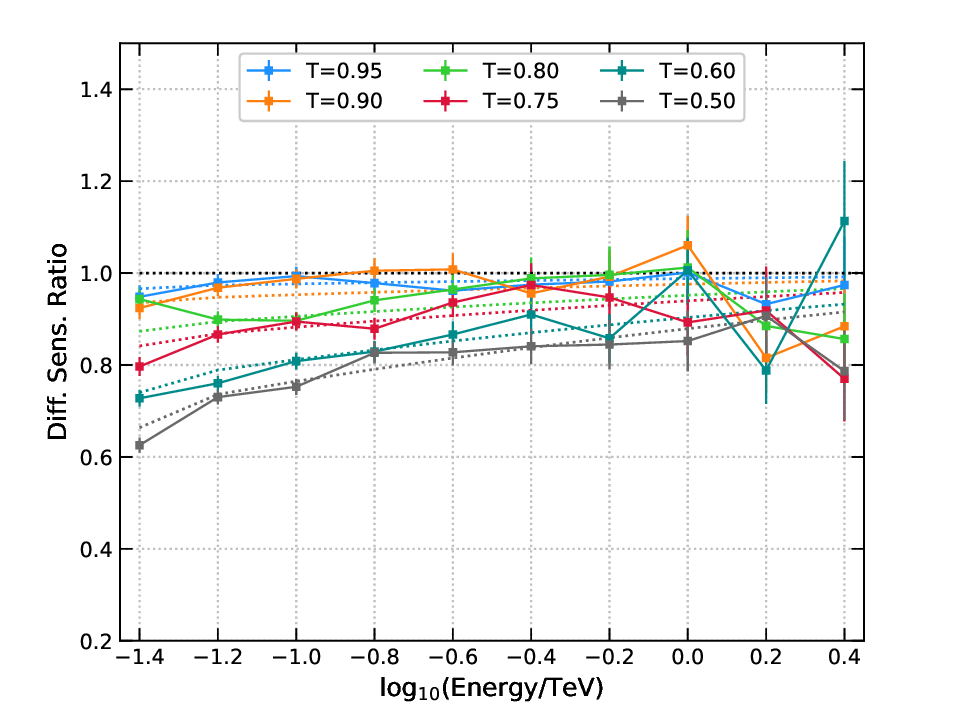}
            }
            \subfigure[13 km a.g.l.] 
            {
                \label{subfig:lab6}
                \includegraphics[width=.45\textwidth]{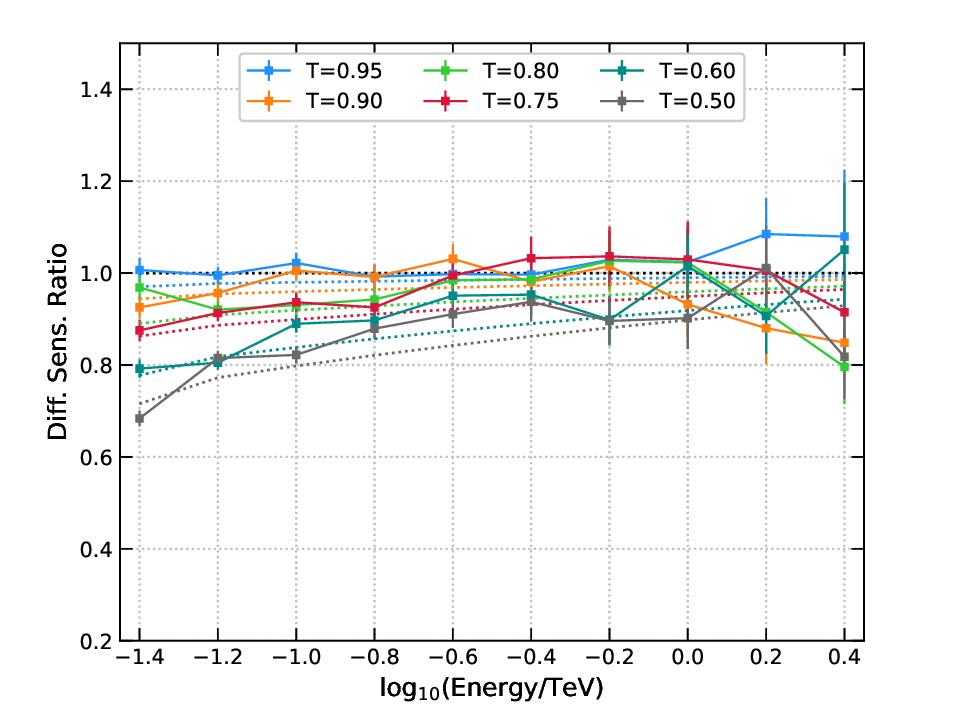} 
            }
            
        \caption{Differential sensitivity ratio $S(T=1)/S(T,H)$ as a function of energy for the subarray of 4 LSTs - a comparison between the cloudless atmosphere and the atmosphere with clouds. Lower values correspond to worse sensitivity. Solid lines represent the results of MC simulations, while dashed lines are the results of the introduced semi-analytical model.}
        \label{fig:subfigsLSTsdiff}
        \end{figure*}

\begin{figure*}[t]
        \centering
            \subfigure[3 km a.g.l.]
            {
                \label{subfig:lab11}
                \includegraphics[width=.45\textwidth]{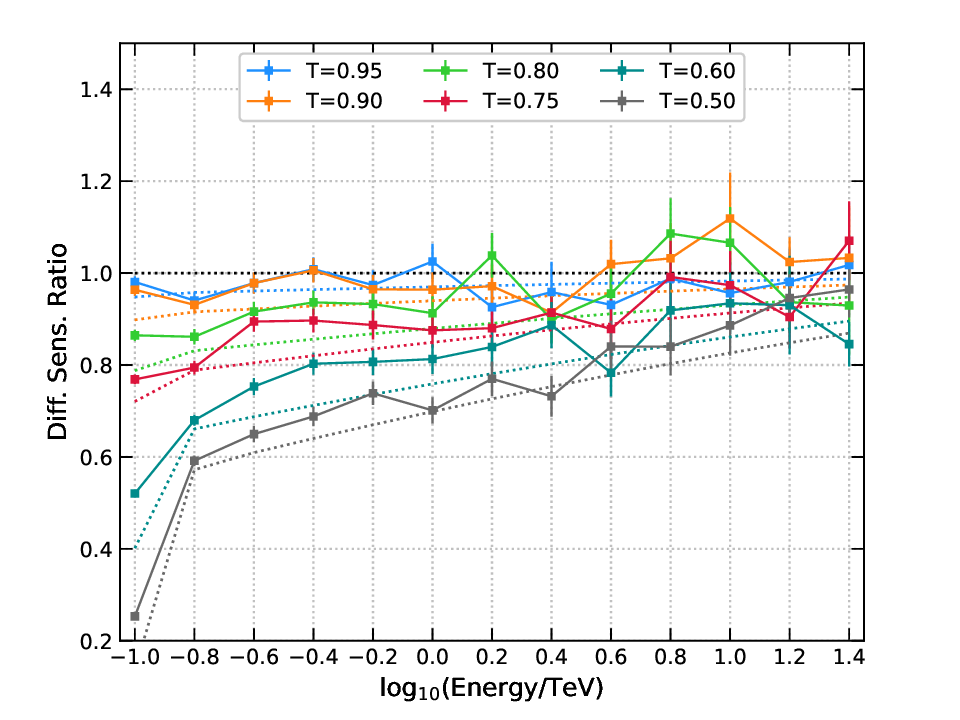} 
            } 
            \subfigure[5 km a.g.l.] 
            {
                \label{subfig:lab22}
                \includegraphics[width=.45\textwidth]{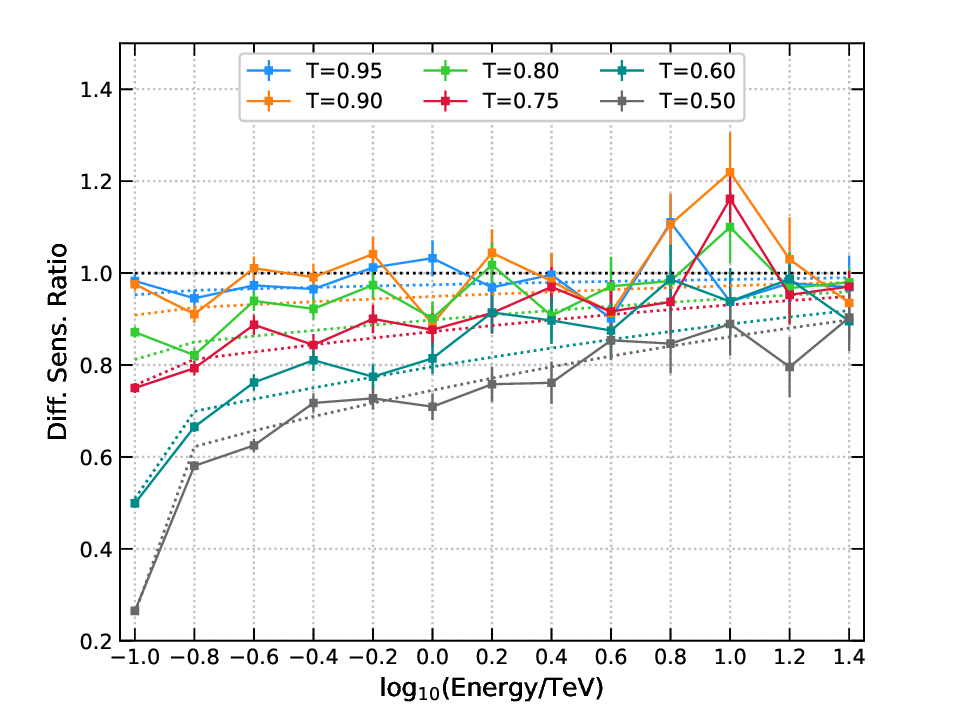} 
            }\\
            \subfigure[7 km a.g.l.] 
            {
                \label{subfig:lab33}
                \includegraphics[width=.45\textwidth]{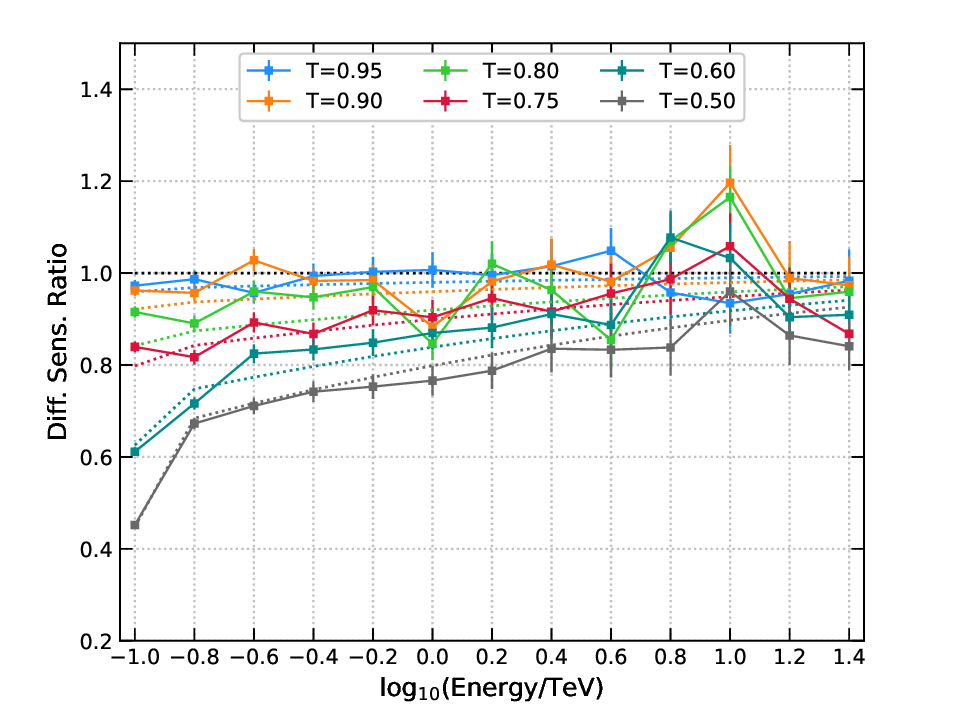} 
            } 
            \subfigure[9 km a.g.l.] 
            {
                \label{subfig:lab44}
                \includegraphics[width=.45\textwidth]{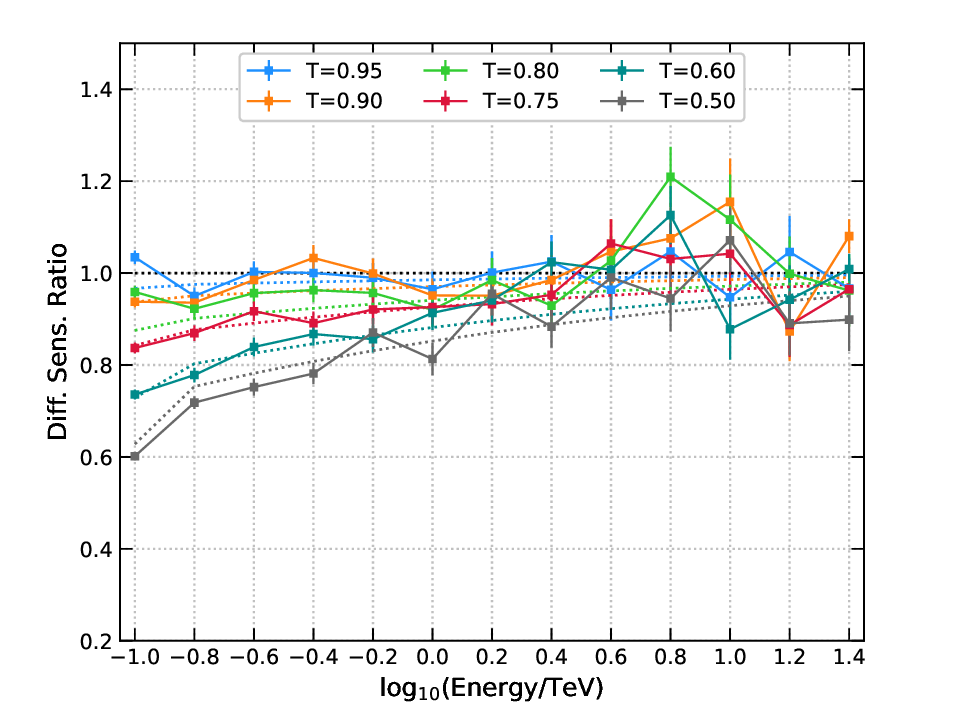} 
            }\\
            \subfigure[11 km a.g.l.]
            {
                \label{subfig:lab55}
                \includegraphics[width=.45\textwidth]{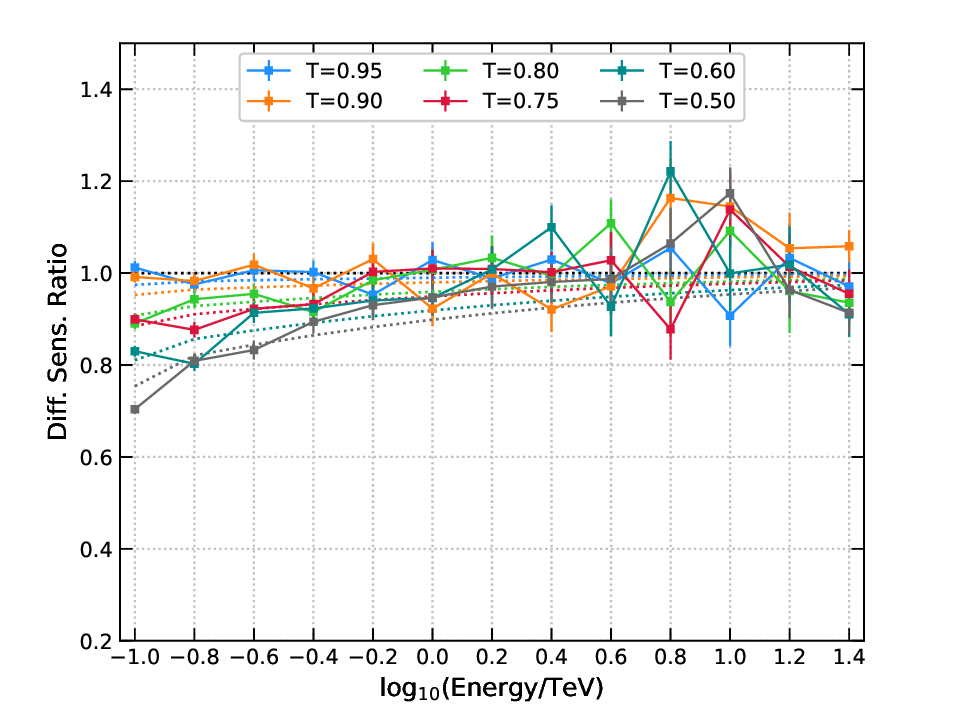} 
            }
            \subfigure[13 km a.g.l.] 
            {
                \label{subfig:lab66}
                \includegraphics[width=.45\textwidth]{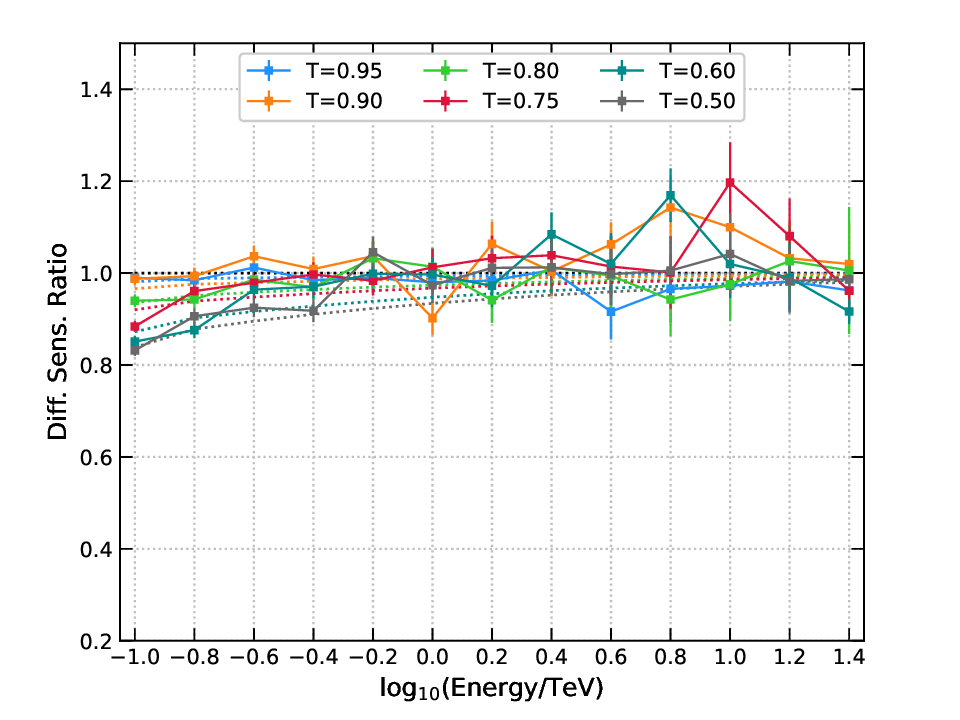} 
            }
        \caption{Differential sensitivity ratio $S(T=1)/S(T,H)$ as a function of energy for the subarray of 15 MSTs - a comparison between the cloudless atmosphere and the atmosphere with clouds. Lower values correspond to worse sensitivity. Solid lines represent the results of MC simulations, while dashed lines are the results of the introduced semi-analytical model.}
        \label{fig:subfigsMSTsdiff}
        \end{figure*}

\begin{figure*}[t]
        \centering
            \subfigure[3 km a.g.l.]
            {
                \label{subfig:lab111}
                \includegraphics[width=.45\textwidth]{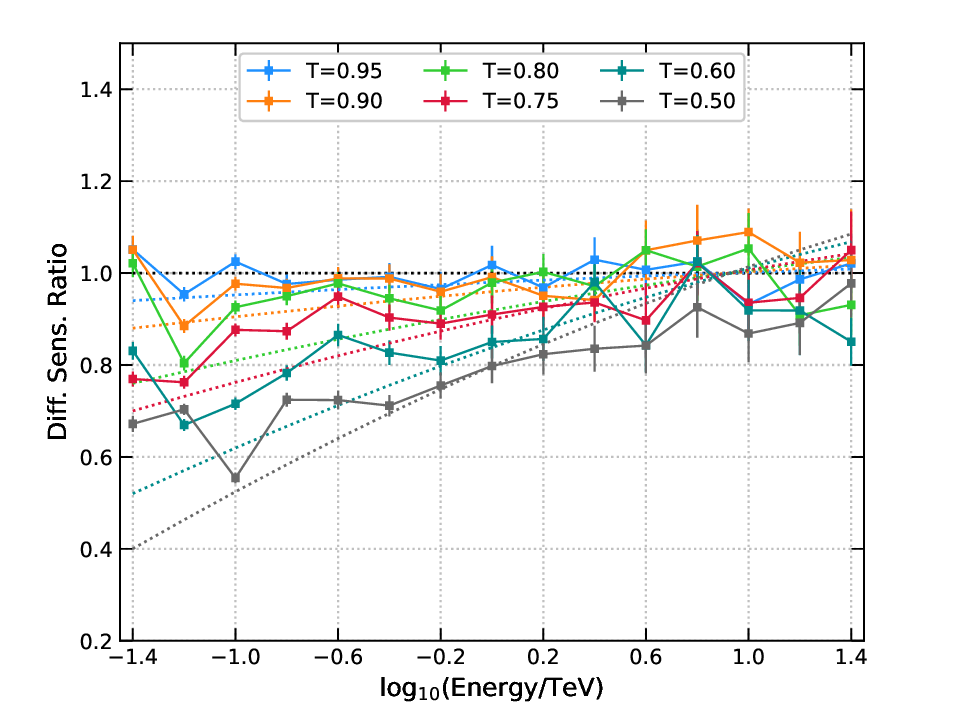} 
            } 
            \subfigure[5 km a.g.l.] 
            {
                \label{subfig:lab222}
                \includegraphics[width=.45\textwidth]{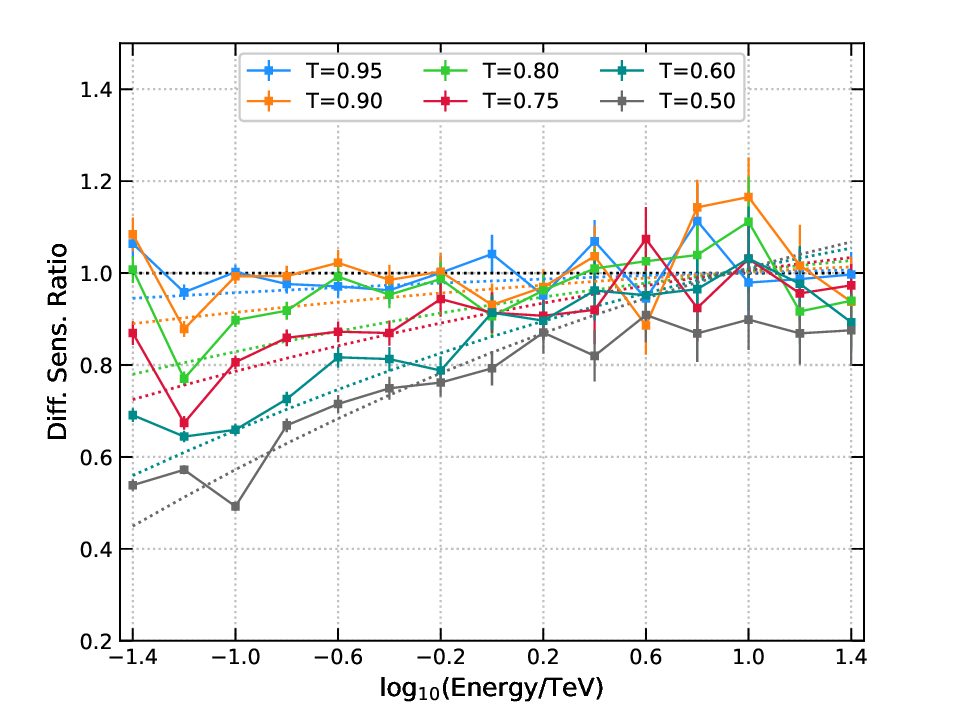} 
            }\\
            \subfigure[7 km a.g.l.] 
            {
                \label{subfig:lab333}
                \includegraphics[width=.45\textwidth]{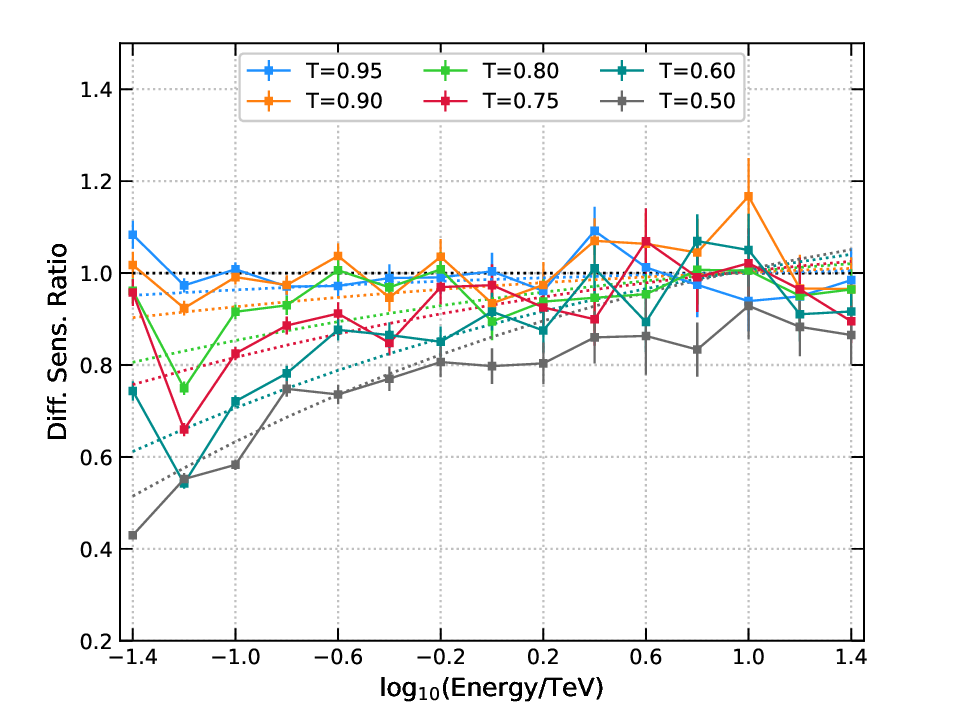} 
            } 
            \subfigure[9 km a.g.l.] 
            {
                \label{subfig:lab444}
                \includegraphics[width=.45\textwidth]{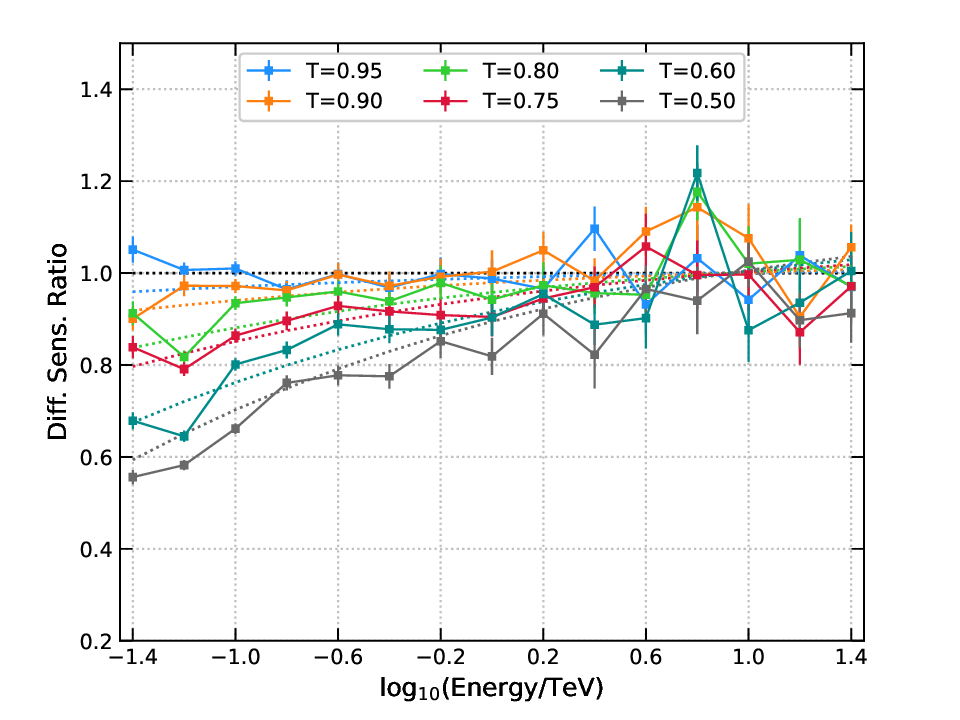} 
            }\\
            \subfigure[11 km a.g.l.] 
            {
                \label{subfig:lab555}
                \includegraphics[width=.45\textwidth]{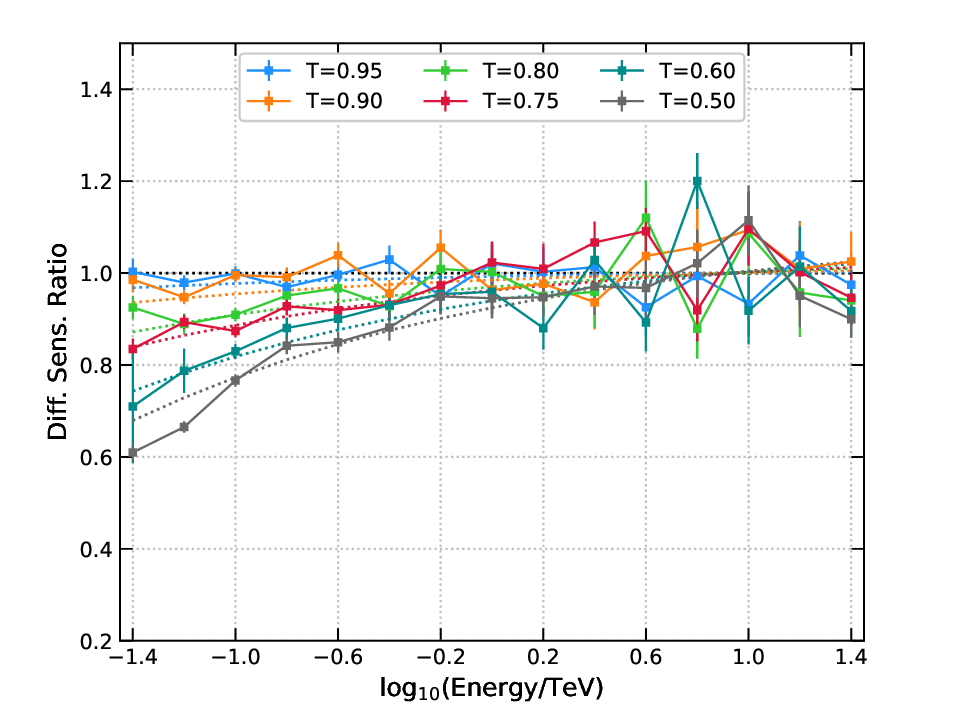} 
            }
            \subfigure[13 km a.g.l.] 
            {
                \label{subfig:lab666}
                \includegraphics[width=.45\textwidth]{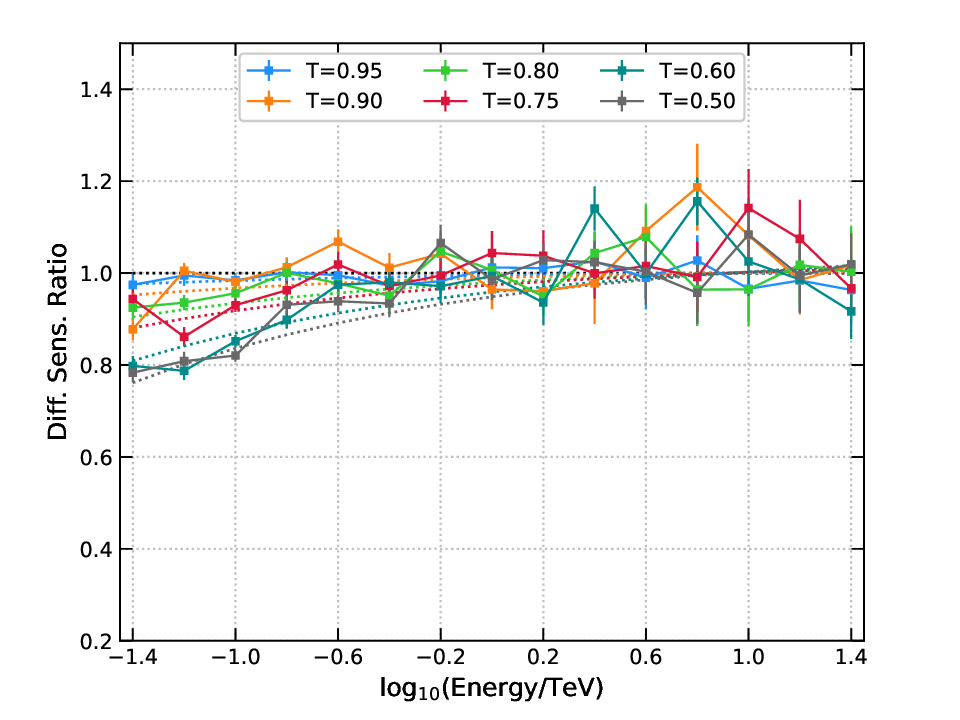} 
            }
        \caption{Differential sensitivity ratio $S(T=1)/S(T,H)$ as a function of energy for the CTA-N array - a comparison between the cloudless atmosphere and the atmosphere with clouds. Lower values correspond to worse sensitivity. Solid lines represent the results of MC simulations, while dashed lines are the results of the introduced semi-analytical model.}
        \label{fig:subfigsCTANdiff}
        \end{figure*}

Having obtained the values of differential sensitivity in a clean atmosphere, $S(T=1)$, it is easy to calculate the ratio of differential sensitivity $S(T=1)/S(T,H)$ using the equation \ref{eq:dsr}. The differential sensitivity ratio obtained with the semi-analytical model is compared with the differential sensitivity ratio obtained from the MC simulations in figures \ref{fig:subfigsLSTsdiff} (4 LSTs), \ref{fig:subfigsMSTsdiff} (15 MSTs) and \ref{fig:subfigsCTANdiff} (CTA-N). 

 \begin{table}[t]
 \centering
\begin{tabular}{@{}llll@{}}
\toprule
         \textbf{Parameter} & \textbf{CTA-N} & \textbf{4 LSTs} & \textbf{15 MSTs}  \\ \midrule
         A & 6.086 & 12.06 & 13.74 \\
         B & 0.1234 & 0.06487 &  0.1361 \\
         C & 0.5559 &  0.6479 & 0.8053\\
         D &  -3.498 & -1.633 &  -1.199 \\
         F & -0.5992 & -0.4141 & -0.2696 \\
         \bottomrule
    \end{tabular}
    \caption{Parameters of the sensitivity degradation factor obtained from the introduced semi-analytical model.}
    \label{tab:model}
\end{table}

In most of the cases, the model is consistent with MC simulations within an accuracy of $\lesssim10\%$.  The exception is the case of very low clouds (3 km a.g.l.), where the model is not able to reproduce the slight gain of the performance with decreasing height of the cloud, in particular for the LST subarray. 
For such low clouds, almost the whole shower seen by telescopes is attenuated by the same factor, resulting in a more uniform image, which is closer to the clear atmosphere case.
Intriguingly, the semi-analytic model prediction for MST subarray is also rather precise even in the case of such low clouds. 
The difference might be related to the larger FoV of MST telescopes, due to which the telescopes could catch the high offset Cherenkov photons generated in the tail of the shower, below the cloud. This would result in uneven absorption of the image even at very low cloud heights. 
However, it should be stressed that the proper description of the threshold effects is very difficult with a simple semi-analytic model.

\clearpage

\clearpage

\begin{sidewaystable}[t!]
\begin{tabular}{@{}cccccccc@{}}
\toprule
\multirow{2}{*}{energy range $\log_{10}(E/$TeV$)$} & \multirow{2}{*}{\textit{T=}1.00 } &  
\multicolumn{2}{c}{3 km a.g.l.}            & \multicolumn{2}{c}{7 km a.g.l.}            & \multicolumn{2}{c}{13 km a.g.l.}           \\ \cmidrule(l){3-8} 
                                          &  &  \textit{T=}0.80 &\textit{T=}0.50  & \textit{T=}0.80 &\textit{T=}0.50  & \textit{T=}0.80 &\textit{T=}0.50  \\  \midrule

-1.4$\pm$0.1 & 5.825$\pm$0.138 & 5.844$\pm$0.151 & 8.881$\pm$0.199 & 6.207$\pm$0.155 & 13.887$\pm$0.205 & 6.452$\pm$0.166 & 7.619$\pm$0.179 \\
-1.2$\pm$0.1 & 1.895$\pm$0.025 & 2.421$\pm$0.042 & 2.767$\pm$0.040  & 2.597$\pm$2.597 & 3.526$\pm$0.074  & 2.080$\pm$0.031  & 2.408$\pm$0.056 \\
-1$\pm$0.1   & 0.978$\pm$0.013 & 1.075$\pm$0.015 & 1.795$\pm$0.035 & 1.087$\pm$1.087 & 1.707$\pm$0.030  & 1.041$\pm$0.015 & 1.213$\pm$0.017 \\
-0.8$\pm$0.1 & 0.611$\pm$0.011 & 0.663$\pm$0.012 & 0.870$\pm$0.016  & 0.677$\pm$0.677 & 0.842$\pm$0.017  & 0.63$\pm$0.013  & 0.677$\pm$0.013 \\
-0.6$\pm$0.1 & 0.424$\pm$0.009 & 0.437$\pm$0.009 & 0.591$\pm$0.015 & 0.425$\pm$0.425 & 0.581$\pm$0.015  & 0.437$\pm$0.011 & 0.455$\pm$0.010 \\
-0.4$\pm$0.1 & 0.286$\pm$0.008 & 0.303$\pm$0.008 & 0.402$\pm$0.012 & 0.295$\pm$0.295 & 0.371$\pm$0.011  & 0.301$\pm$0.009 & 0.306$\pm$0.009 \\
-0.2$\pm$0.1 & 0.197$\pm$0.006 & 0.230$\pm$0.008  & 0.280$\pm$0.009  & 0.210$\pm$0.219 & 0.262$\pm$0.011  & 0.202$\pm$0.007 & 0.199$\pm$0.006 \\
0$\pm$0.1    & 0.164$\pm$0.007 & 0.162$\pm$0.007 & 0.199$\pm$0.008 & 0.177$\pm$0.177 & 0.199$\pm$0.009  & 0.158$\pm$0.006 & 0.161$\pm$0.007 \\
0.2$\pm$0.1  & 0.133$\pm$0.006 & 0.131$\pm$0.004 & 0.159$\pm$0.008 & 0.140$\pm$0.140 & 0.163$\pm$0.008  & 0.138$\pm$0.005 & 0.128$\pm$0.004 \\
0.4$\pm$0.1  & 0.129$\pm$0.007 & 0.128$\pm$0.005 & 0.149$\pm$0.008 & 0.131$\pm$0.131 & 0.144$\pm$0.008  & 0.119$\pm$0.004 & 0.121$\pm$0.004 \\
0.6$\pm$0.1  & 0.123$\pm$0.005 & 0.129$\pm$0.004 & 0.161$\pm$0.010  & 0.142$\pm$0.142 & 0.157$\pm$0.014  & 0.126$\pm$0.007 & 0.135$\pm$0.009 \\
0.8$\pm$0.1  & 0.171$\pm$0.011 & 0.165$\pm$0.011 & 0.181$\pm$0.011 & 0.166$\pm$0.166 & 0.201$\pm$0.013  & 0.174$\pm$0.013 & 0.175$\pm$0.011 \\
1$\pm$0.1    & 0.199$\pm$0.009 & 0.212$\pm$0.014 & 0.257$\pm$0.016 & 0.222$\pm$0.222 & 0.240$\pm$0.017  & 0.231$\pm$0.018 & 0.206$\pm$0.012 \\
1.2$\pm$0.1  & 0.306$\pm$0.032 & 0.331$\pm$0.020 & 0.337$\pm$0.022 & 0.316$\pm$0.316 & 0.340$\pm$0.020  & 0.295$\pm$0.020  & 0.302$\pm$0.021 \\
1.4$\pm$0.1  & 0.424$\pm$0.042 & 0.487$\pm$0.028 & 0.463$\pm$0.029 & 0.470$\pm$0.470 & 0.523$\pm$0.032  & 0.452$\pm$0.040  & 0.444$\pm$0.023 \\ \hline
\end{tabular}

\caption{Comparative view of the influence of the most dominant clouds (3 km a.g.l.), the most often clouds at La Palma (7 km a.g.l.), and the least dominant clouds (13 km a.g.l.) on the differential sensitivity of CTA-N in units of $10^{-12}$ erg cm$^{-2}$ s$^{-1}$.}
\label{tab:diffsens}
\end{sidewaystable}

\clearpage

\begin{figure*}[t]
        \centering
                    \subfigure[Effective Area, 7 km a.g.l.] 
            {
                \includegraphics[width=.45\textwidth]{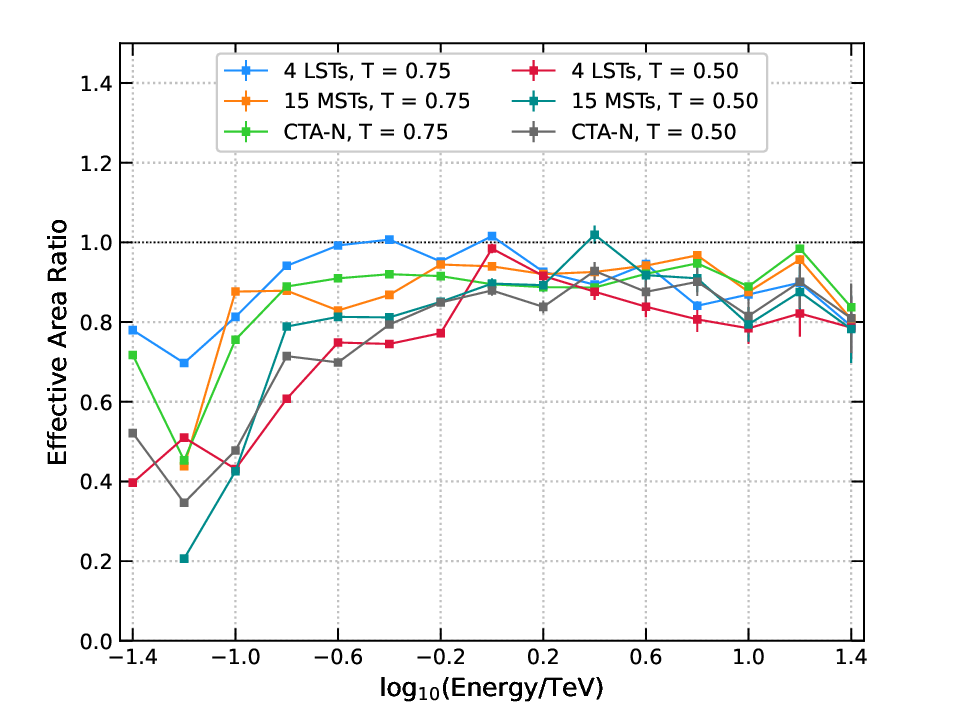} 
            } 
            \subfigure[Effective Area, 9 km a.g.l.] 
            {
                \includegraphics[width=.45\textwidth]{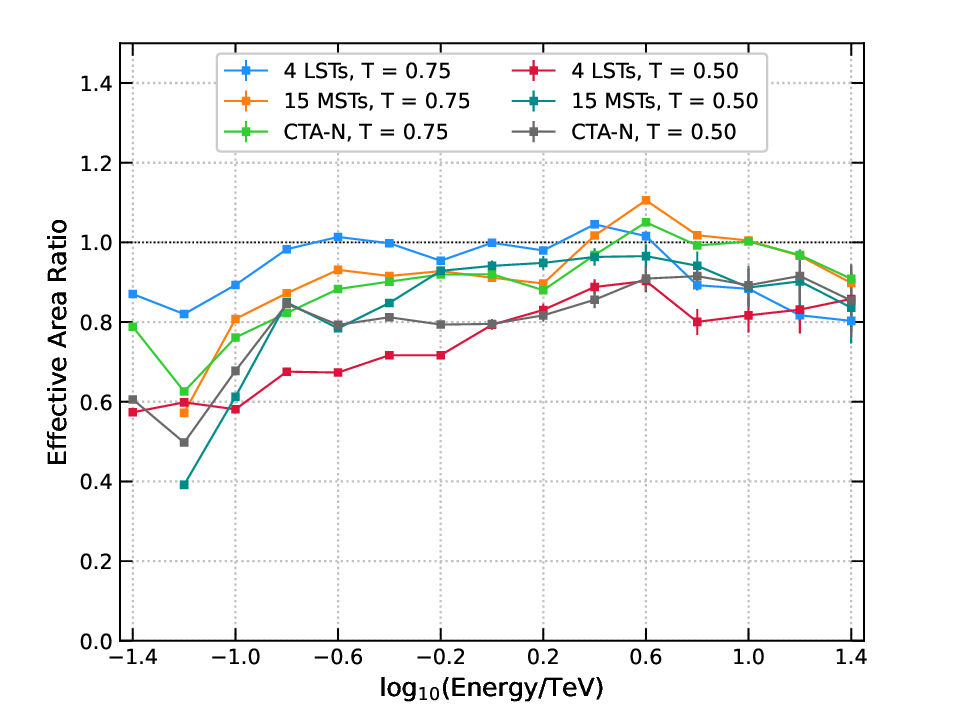} 
            } 
            \subfigure[Diff. Sens., 7 km a.g.l.] 
            {
                \includegraphics[width=.45\textwidth]{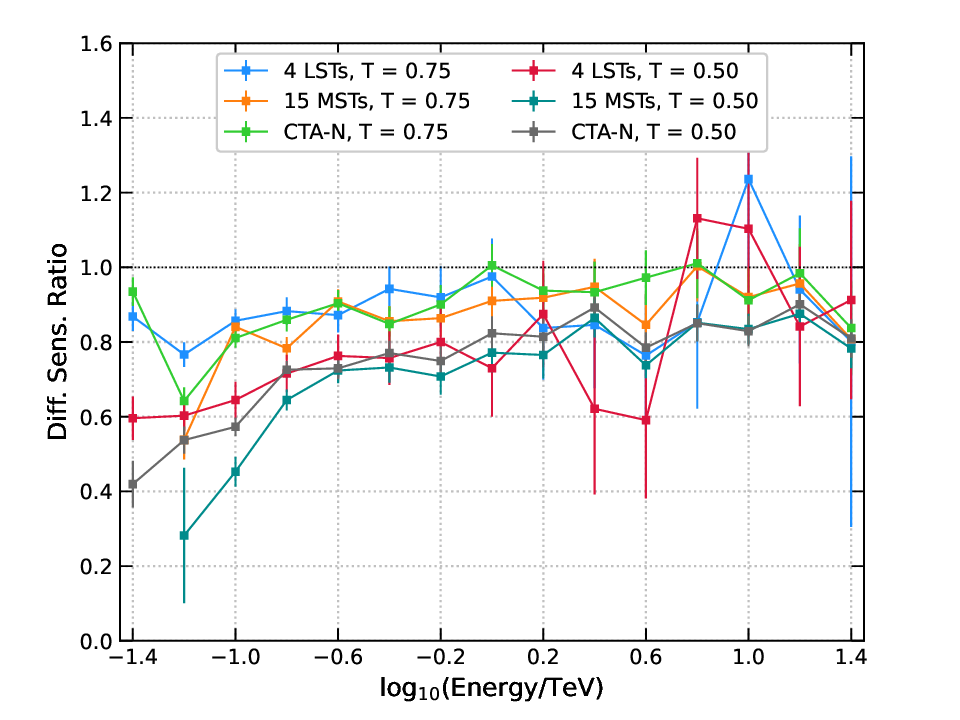} 
            } 
            \subfigure[Diff. Sens., 9 km a.g.l.] 
            {
                \includegraphics[width=.45\textwidth]{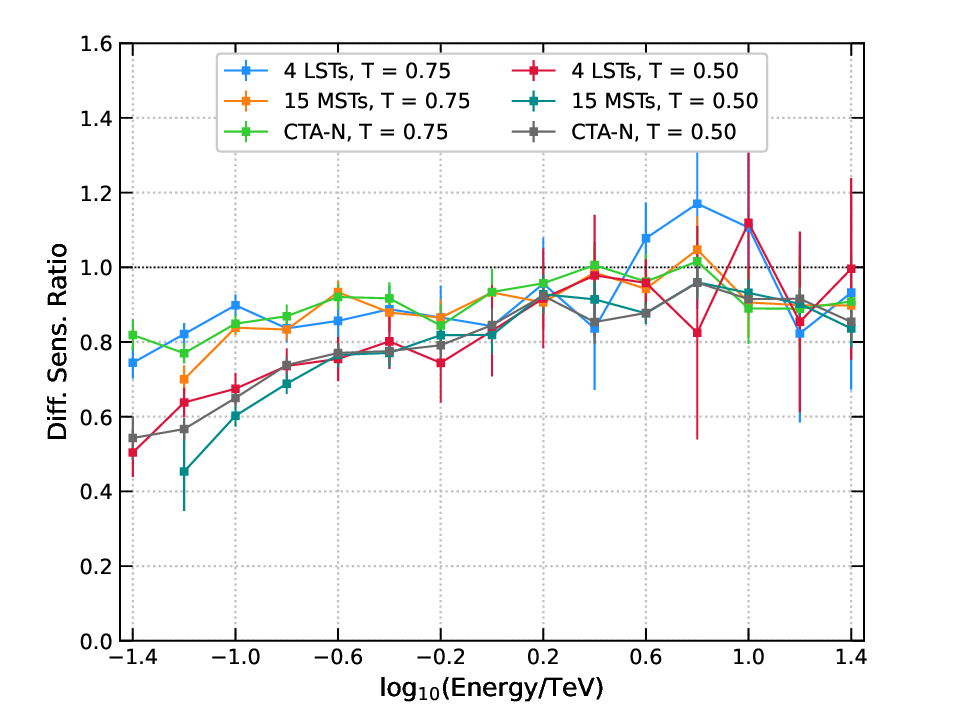} 
            } 

\caption{Comparative view of the influence of most frequent clouds over La Palma (at 7 and 9 km a.g.l.) on the differential sensitivity and effective area of individual subarrays and full CTA-N array. Lower values correspond to worse sensitivity and a smaller effective area.}
\label{fig:comparison_ALL}
\end{figure*}

Figure \ref{fig:comparison_ALL} shows a comparative view of the influence of most frequent clouds over La Palma (7 and 9 km a.g.l.), for $T=0.50$ and $T=0.75$, on the differential sensitivity (bottom panels) and effective area (upper panels) of CTA and its corresponding subarrays. Lower clouds at 7 km a.g.l. have a stronger impact on the effective area compared to higher clouds, reducing it by up to $\approx$ 45\%, $\approx$ 80\% and $\approx$ 60\% at 60 GeV, respectively, for subarrays of 4 LSTs, 15 MSTs, and CTA-N, depending on the transmission.  Above 1 TeV, the effective area is reduced by no more than 20\%. As expected, at lower energies clouds have a dominant influence on MSTs, and at higher energies on LSTs.  

The effective area in the presence of clouds is reduced primarily because of a decreased gamma-ray detection efficiency, which depends on the number of detected events. Whether the event is detected or not depends on the relationship between the height of the shower maximum and the cloud altitude, or more precisely on the fraction of photons produced above the cloud. Low and middle clouds are on average below or within the shower maximum and influence a larger number of Cherenkov photons compared to higher clouds. Consequently, the sensitivity is also proportionally worsened, especially in the lower energy bins.

\clearpage
\subsection{Energy resolution}
Energy resolution describes how accurately the instrument can determine the real energy of an event. The energy resolution, $\delta r$ is defined as half the width of the interval in $r=E_\mathrm{est}/E_\mathrm{true}$ symmetric around 1 that contains 68\% of the distribution, namely:
\begin{equation*}
    \int_{1-\delta r}^{1+\delta r}\frac{\mathrm{d}N}{\mathrm{d}r}\mathrm{d}r=0.68.
\end{equation*} 
Only events that passed the quality cuts used to calculate the respective differential sensitivity were used in the calculation of the energy resolution. Five energy bins in $E_\mathrm{est}$ per decade were selected for the estimation of the energy resolution.

\begin{figure*}[t]\centerline{}
        \centering
            \subfigure[3 km a.g.l.] 
            {
                \includegraphics[width=.45\textwidth]{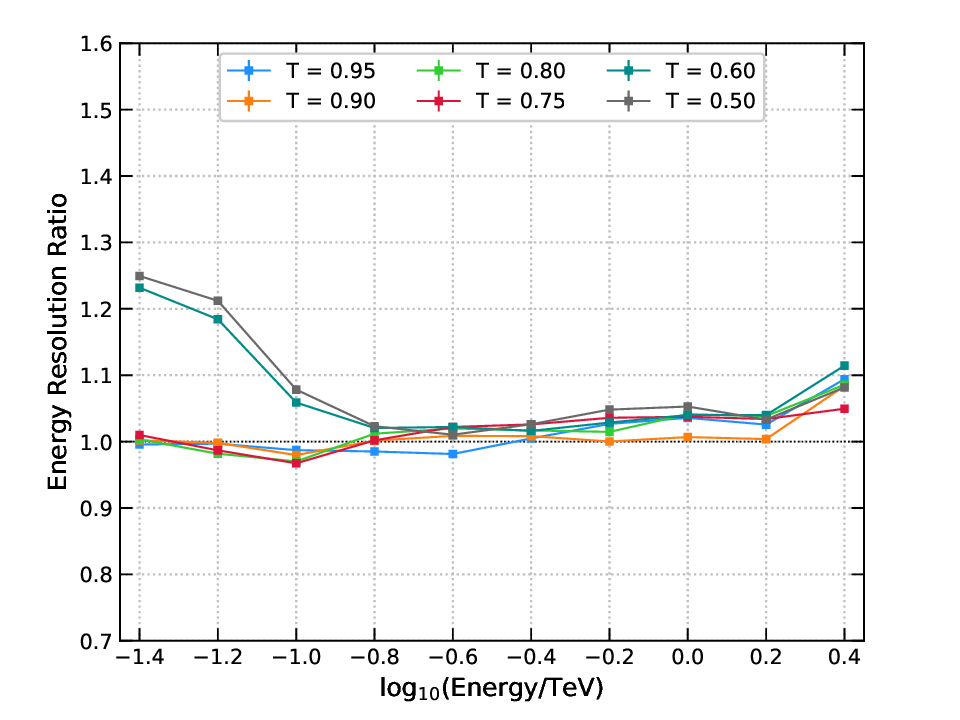} 
            } 
            \subfigure[5 km a.g.l.] 
            {
                \includegraphics[width=.45\textwidth]{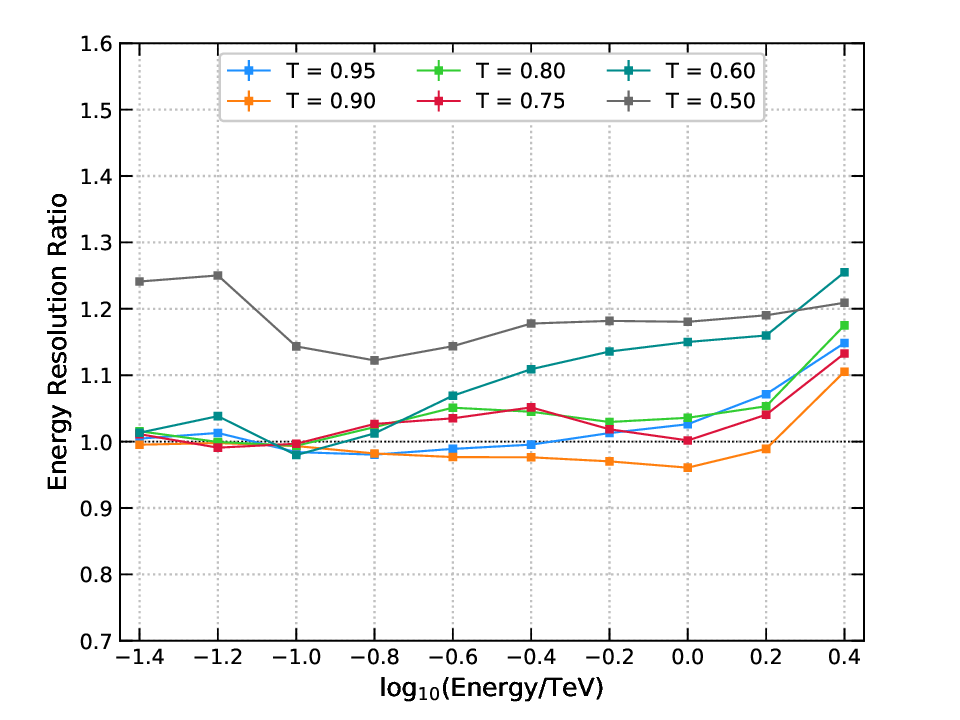} 
            }\\
            \subfigure[7 km a.g.l.] 
            {
                \includegraphics[width=.45\textwidth]{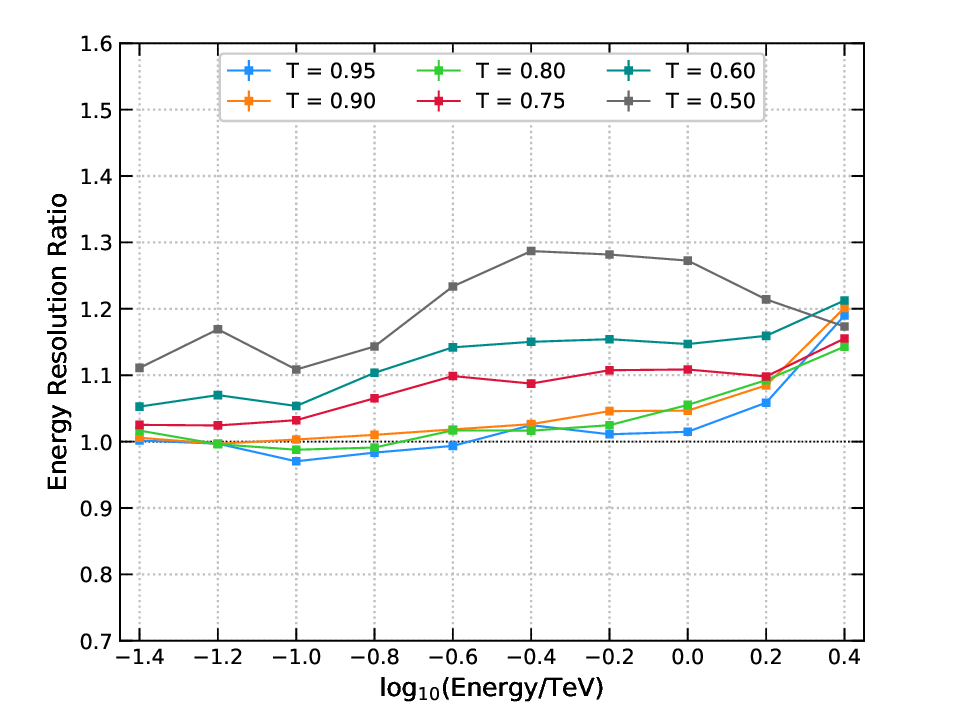}
            } 
            \subfigure[9 km a.g.l.] 
            {
                \includegraphics[width=.45\textwidth]{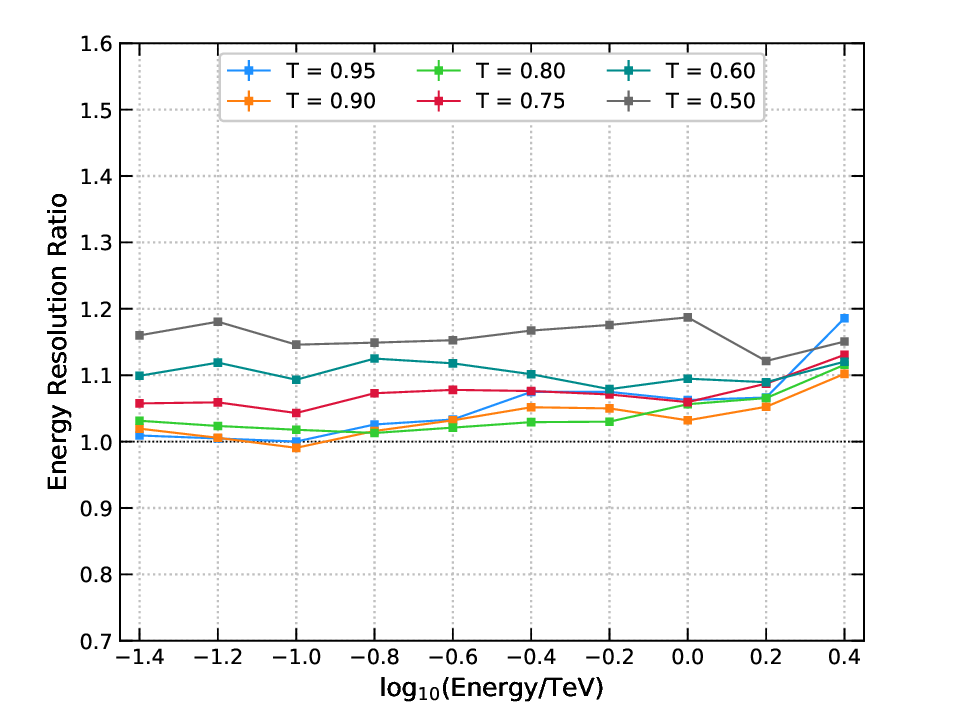} 
            }\\
            
            \subfigure[11 km a.g.l.] 
            {
                \includegraphics[width=.45\textwidth]{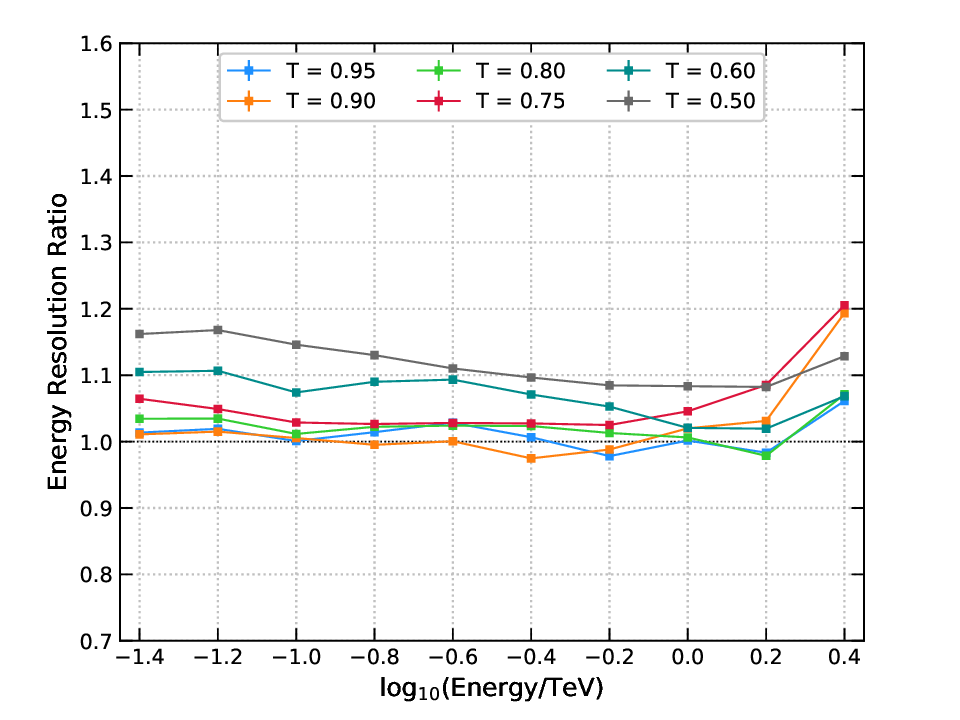}
            }
            \subfigure[13 km a.g.l.] 
            {
                \includegraphics[width=.45\textwidth]{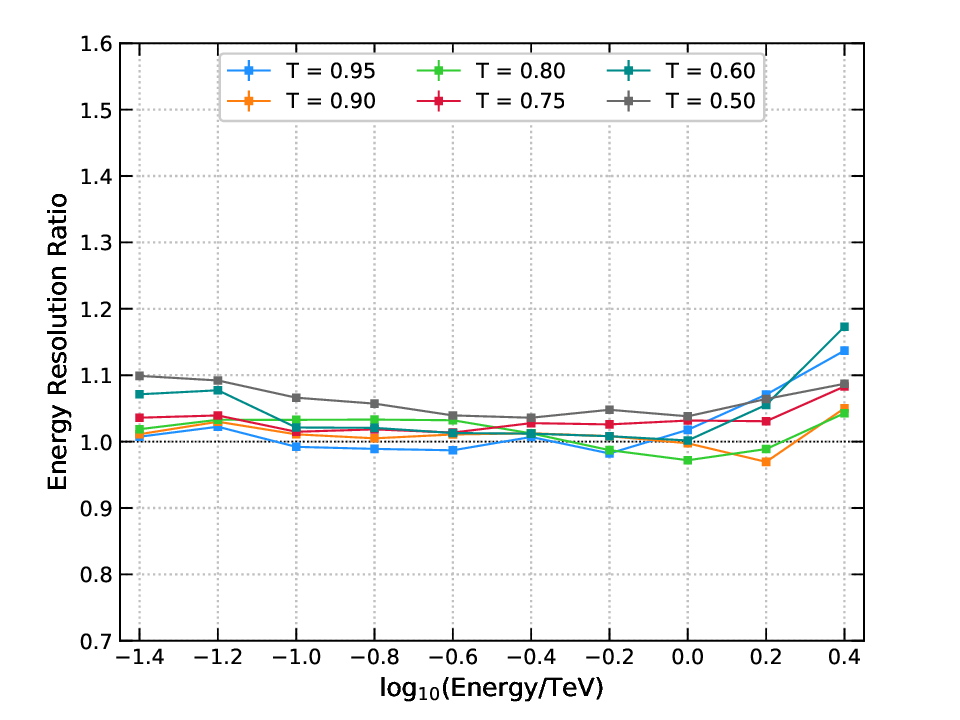} 
            }
            
        \caption{Energy resolution ratio as a function of energy for the subarray of 4 LSTs - a comparison between the cloudless atmosphere and the atmosphere with clouds. The lower value corresponds to better resolution.}
        \label{fig:subfigsLSTsEnRes}
        \end{figure*}

\begin{figure*}[t]
        \centering
            \subfigure[3 km a.g.l.]
            {
                \includegraphics[width=.45\textwidth]{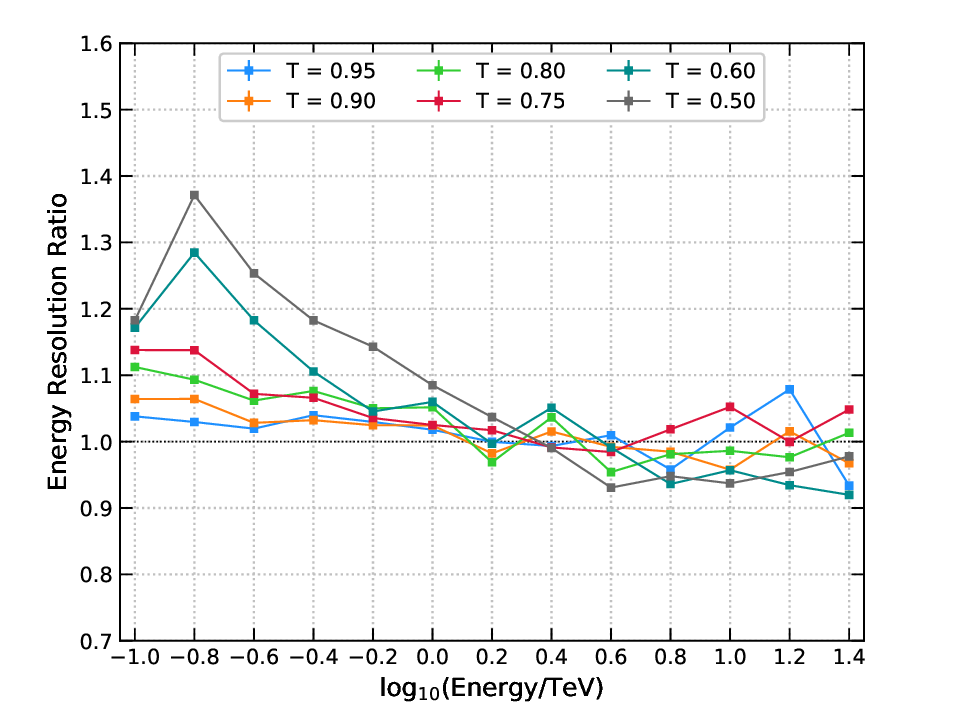}
            } 
            \subfigure[5 km a.g.l.] 
            {
                \includegraphics[width=.45\textwidth]{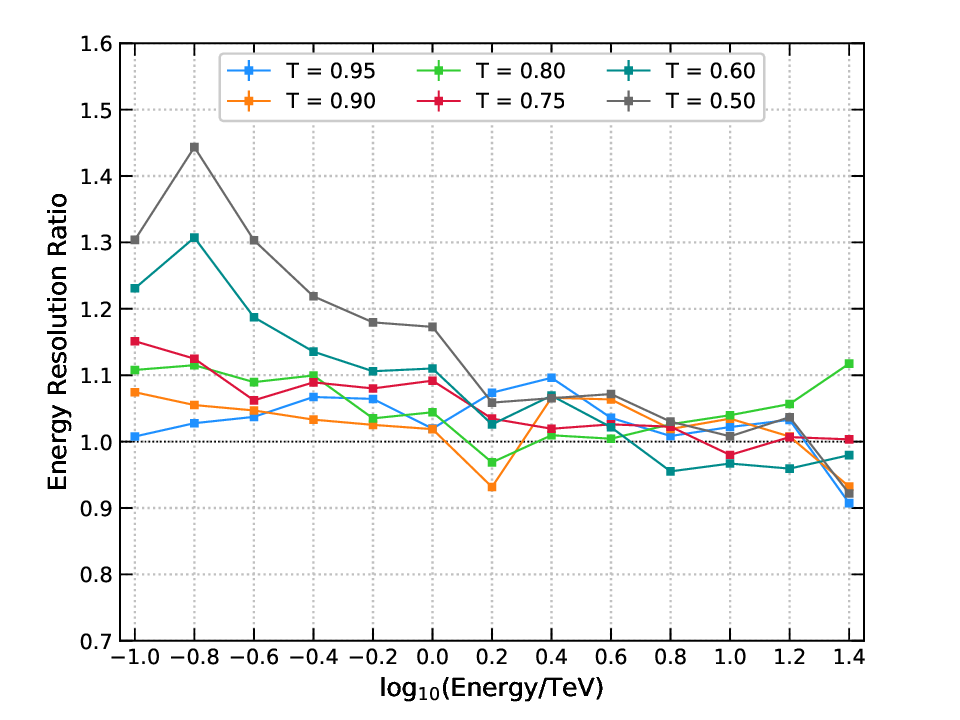}
            }\\
            \subfigure[7 km a.g.l.] 
            {
                \includegraphics[width=.45\textwidth]{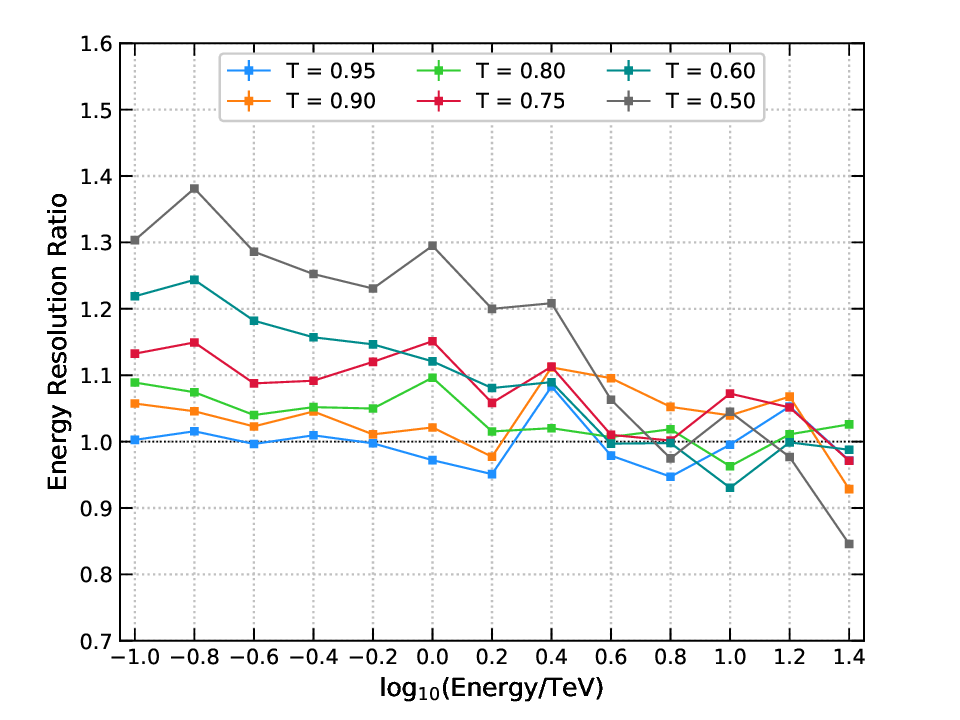} 
            } 
            \subfigure[9 km a.g.l.] 
            {
                \includegraphics[width=.45\textwidth]{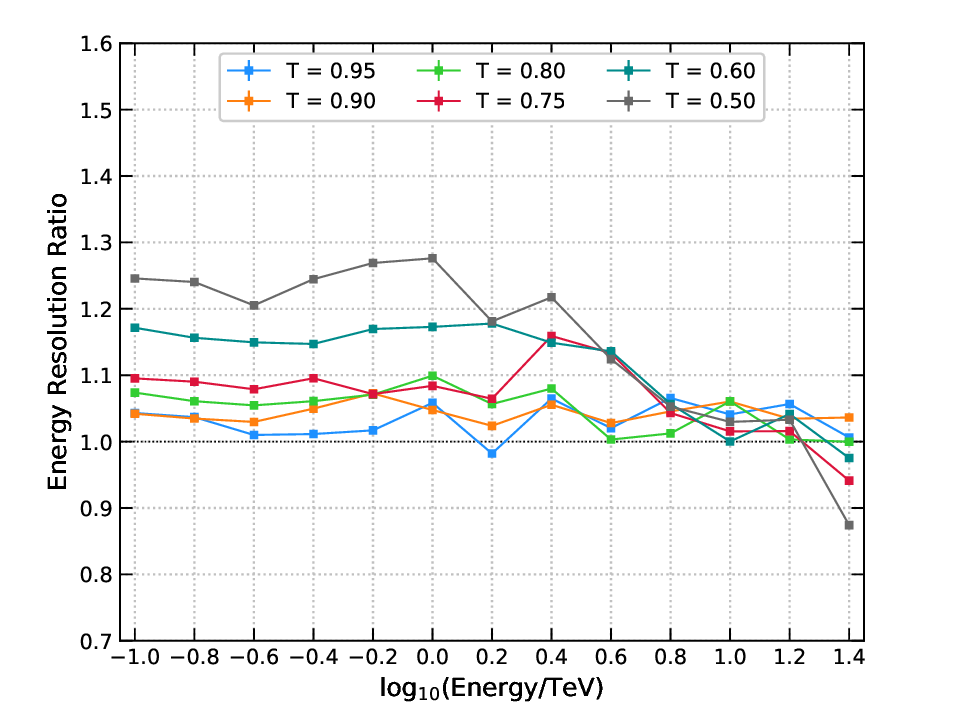} 
            }\\
            \subfigure[11 km a.g.l.] 
            {
                \includegraphics[width=.45\textwidth]{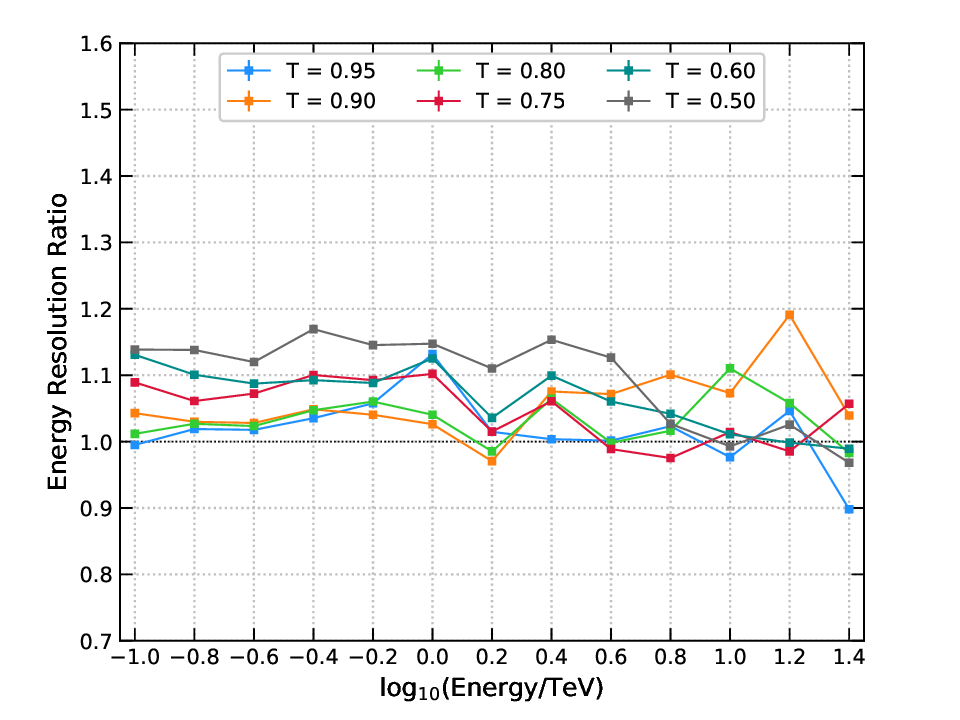} 
            }
            \subfigure[13 km a.g.l.] 
            {
                \includegraphics[width=.45\textwidth]{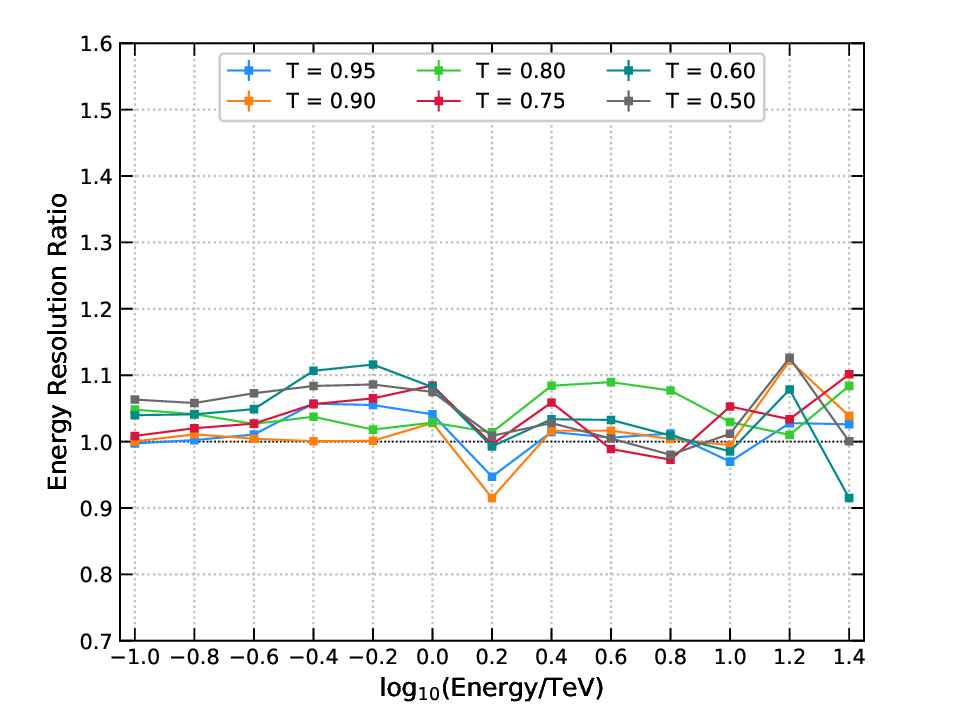}
            }
        \caption{Energy resolution ratio as a function of energy for the subarray of 15 MSTs - a comparison between the cloudless atmosphere and the atmosphere with clouds. The lower value corresponds to better resolution.}
        \label{fig:subfigsMSTsEnRes}
        \end{figure*}

In the presence of low and dense clouds ($\leq9$ km a.s.l., $T\leq 0.75$), the energy resolution for the layouts of 4 LSTs (Figure \ref{fig:subfigsLSTsEnRes}) is notably worse compared to the performance in cloudless conditions over the whole energy range, with a relative difference of up to $30\%$. With increasing altitude, the relative differences are $\lesssim$ 10\%, while clouds at 13 km a.g.l. have a dominant impact only in case of very low transmissions ($T\leq 0.60$), and at energies below 200 GeV. In the presence of clouds at 3 km a.g.l., the energy resolution is worse by, depending on the transmission, between 5\% and 20\% at the energy of $\approx 40$ GeV, and by $\leq$ 10\% at 2.5 TeV, compared to the cloudless conditions. The subarray of 15 MSTs (figure \ref{fig:subfigsMSTsEnRes}) suffers from similar trends.  For the cloud transmissions $\geq$ 0.60 and energies above 2.5 TeV, the relative difference in energy resolution is less than 15\%. For example, the energy resolution in the presence of clouds with $T=0.50$ is worse by 10\% to 40\% at an energy of 250 GeV (depending on the cloud altitude), and by $\approx$ 10\% at 25 TeV. 

\begin{figure*}[t]
        \centering
            \subfigure[3 km a.g.l.]
            {
                \includegraphics[width=.45\textwidth]{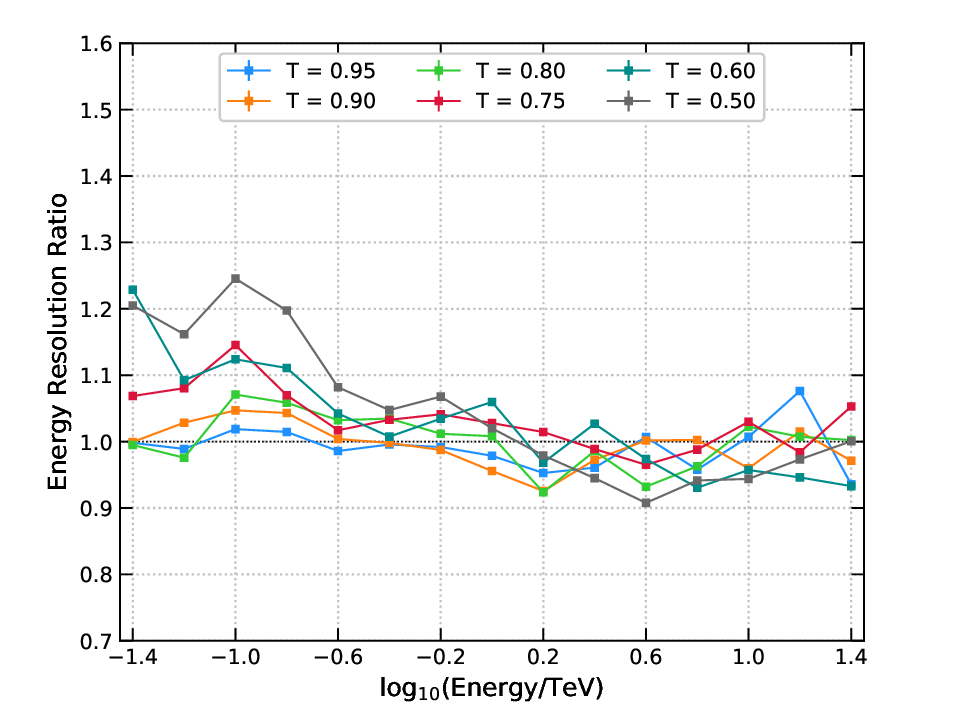}
            } 
            \subfigure[5 km a.g.l.] 
            {
                \includegraphics[width=.45\textwidth]{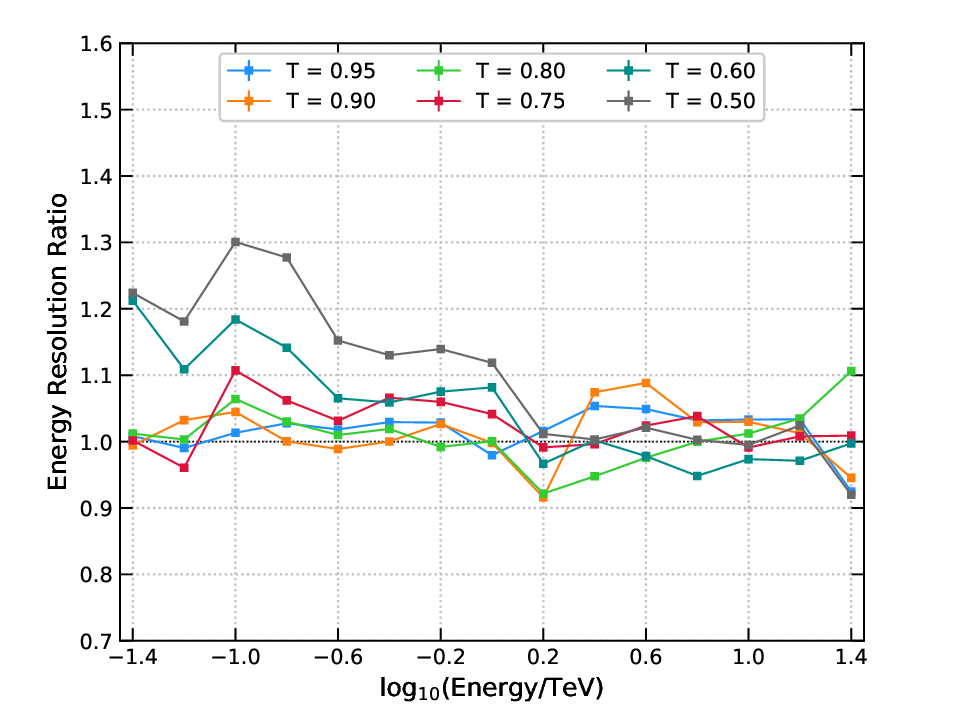}
            }\\
            \subfigure[7 km a.g.l.] 
            {
                \includegraphics[width=.45\textwidth]{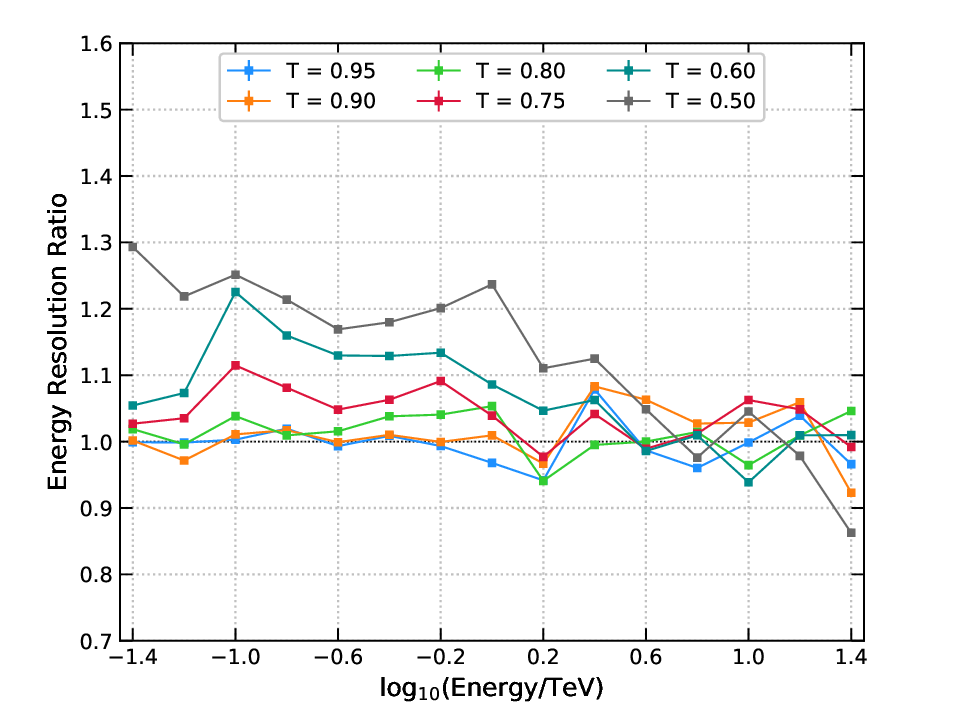} 
            } 
            \subfigure[9 km a.g.l.] 
            {
                \includegraphics[width=.45\textwidth]{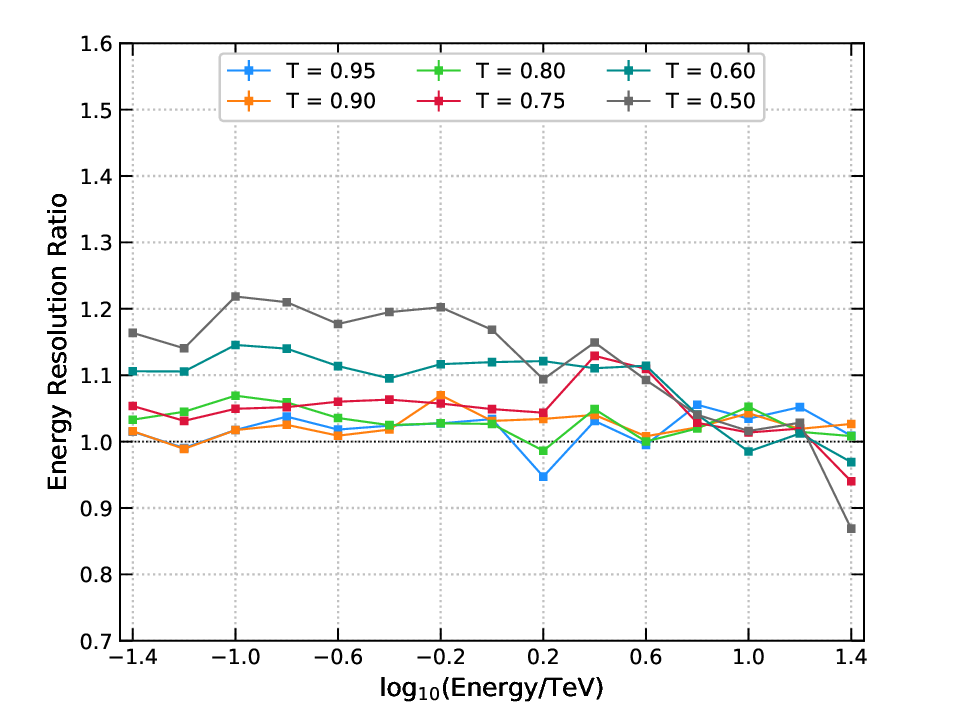} 
            }\\
            \subfigure[11 km a.g.l.] 
            {
                \includegraphics[width=.45\textwidth]{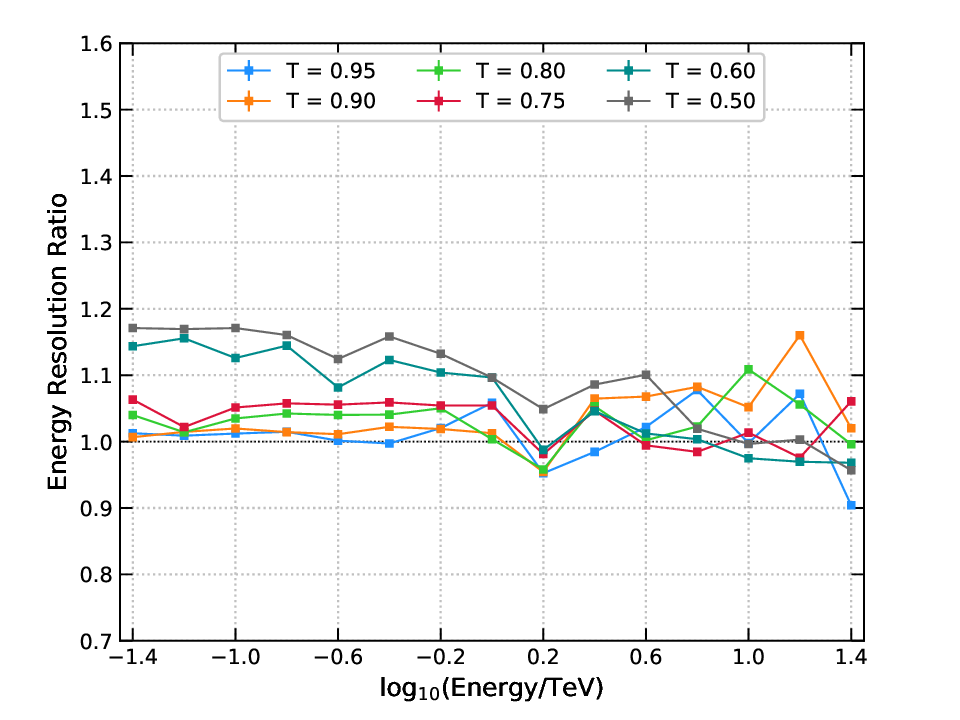} 
            }
            \subfigure[13 km a.g.l.] 
            {
                \includegraphics[width=.45\textwidth]{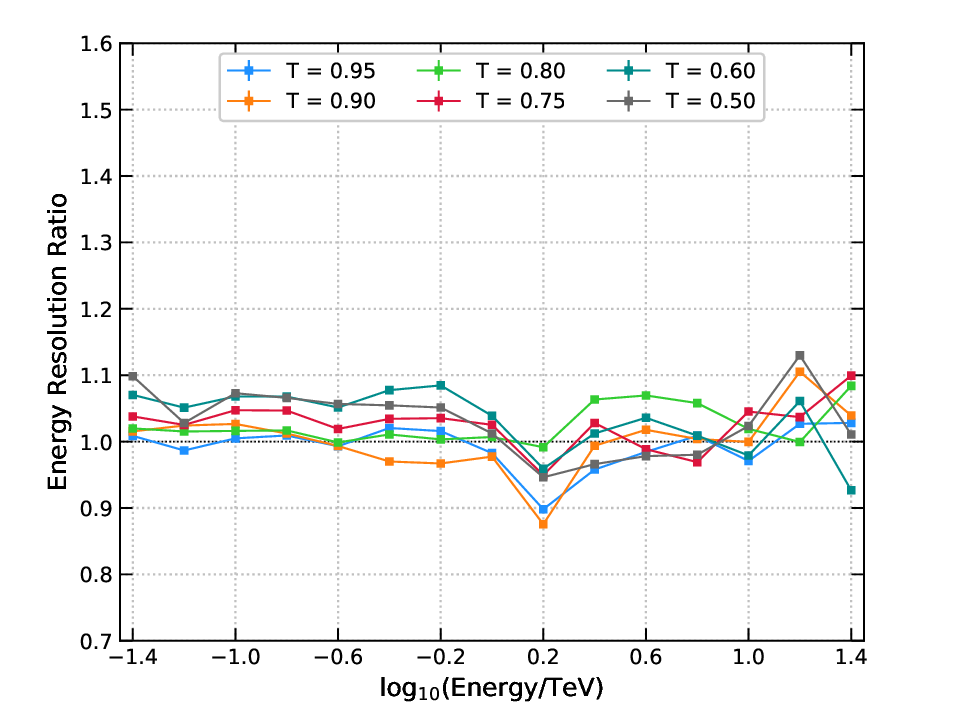}
            }
        \caption{Energy resolution ratio as a function of energy for the CTA-N array - a comparison between the cloudless atmosphere and the atmosphere with clouds. The lower value corresponds to better resolution.}
        \label{fig:subfigsCTAEnRes}
        \end{figure*}

Although the energy resolution is improved over the entire energy range for CTA-N (figure \ref{fig:subfigsCTAEnRes}) compared to the energy resolution of individual subarrays, the dominant effect of low and dense clouds is still evident below 2.5 TeV. The energy resolution degradation is stable across the entire energy range in the presence of high clouds ($\geq 9$ km a.g.l.), while for the low clouds, the effects are slightly higher at the energy threshold. The highly degraded resolution of the telescope at low energies is not unexpected considering that at low energies the number of emitted photons is small, resulting in high fluctuations in the obtained values.

\subsection{Angular resolution}
The angular resolution is a measure of the accuracy with which the direction of the source can be determined, and it is obtained from the distribution of the parameter $\theta$. This parameter indicates the angular distance between the reconstructed and the real source direction. The angular resolution represents the angle at which 68\% of the reconstructed gamma-ray events in the given energy bin fall relative to their true direction.

\begin{figure*}[t]
        \centering
            \subfigure[3 km a.g.l.] 
            {
                \includegraphics[width=.45\textwidth]{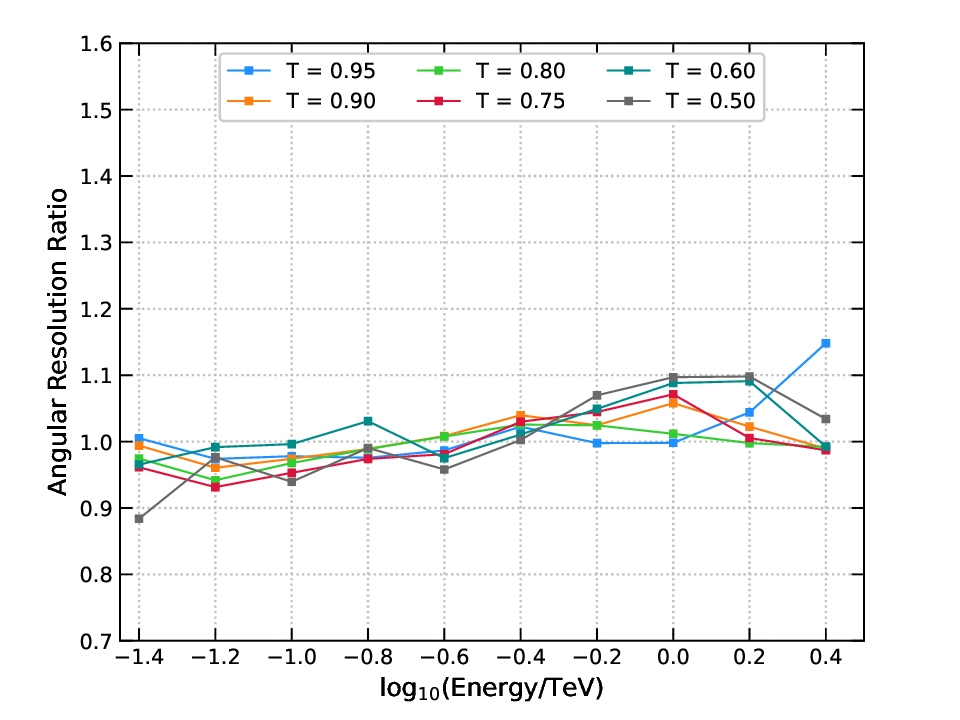} 
            } 
            \subfigure[5 km a.g.l.] 
            {
                \includegraphics[width=.45\textwidth]{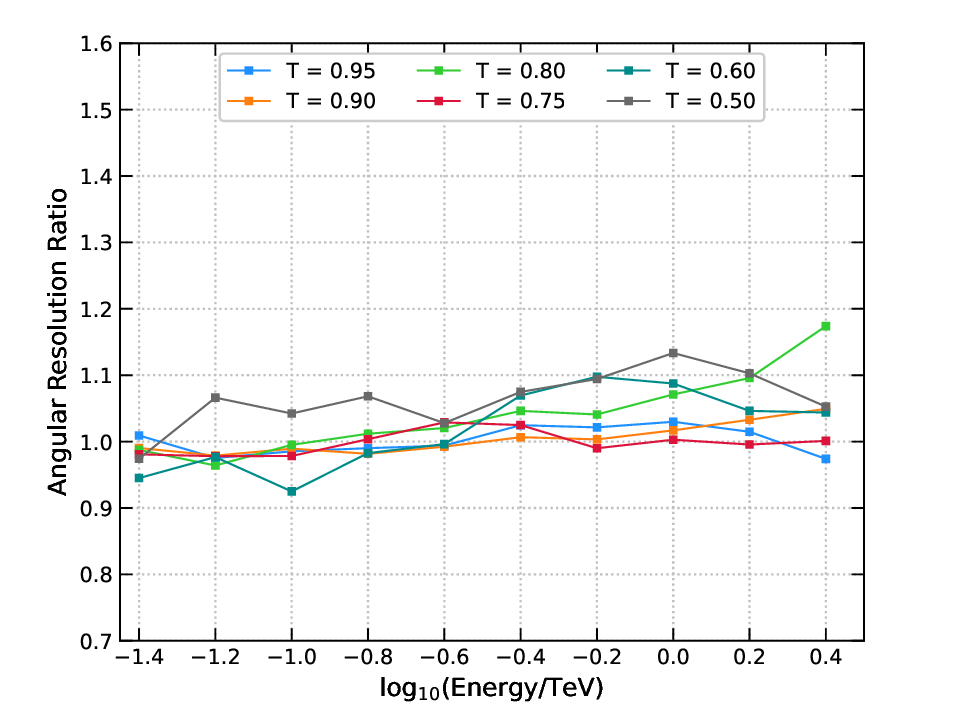} 
            }\\
            \subfigure[7 km a.g.l.] 
            {
                \includegraphics[width=.45\textwidth]{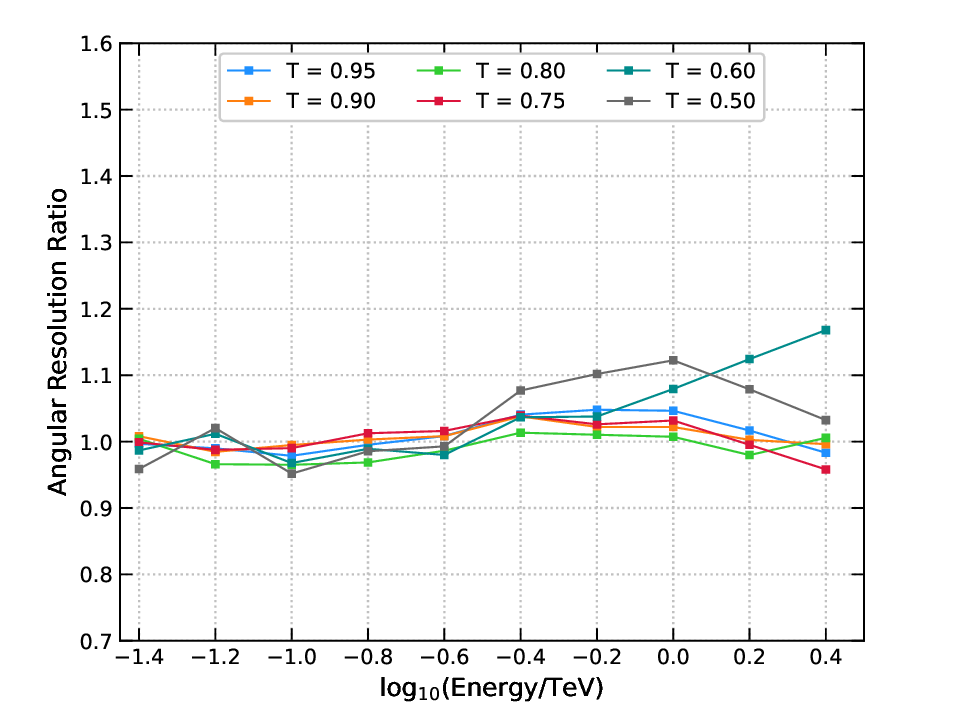}
            } 
            \subfigure[9 km a.g.l.] 
            {
                \includegraphics[width=.45\textwidth]{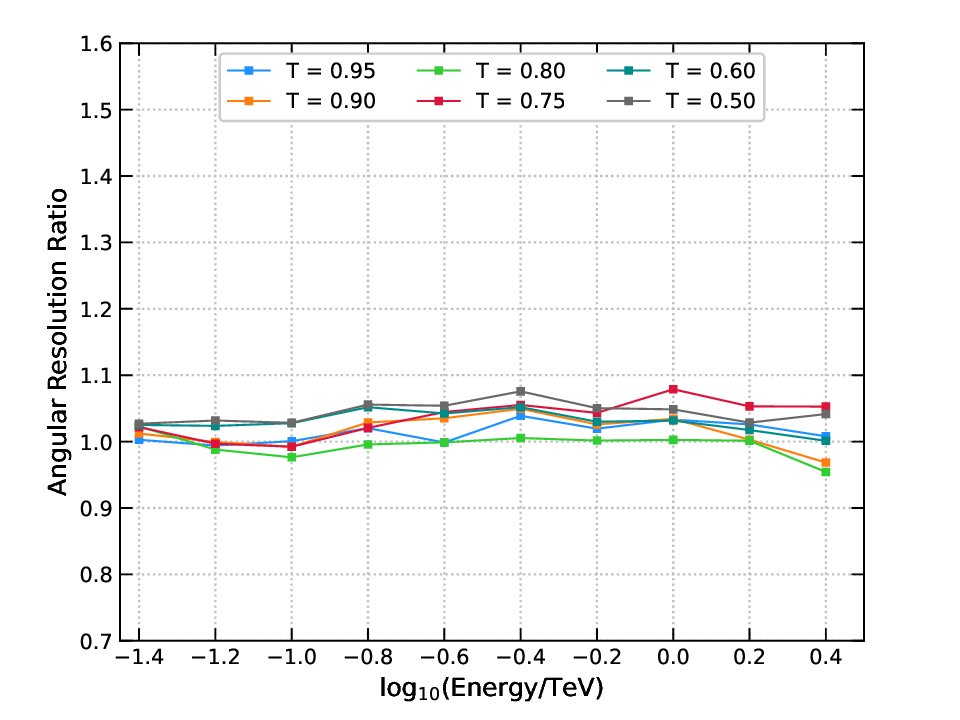} 
            }\\
            
            \subfigure[11 km a.g.l.] 
            {
                \includegraphics[width=.45\textwidth]{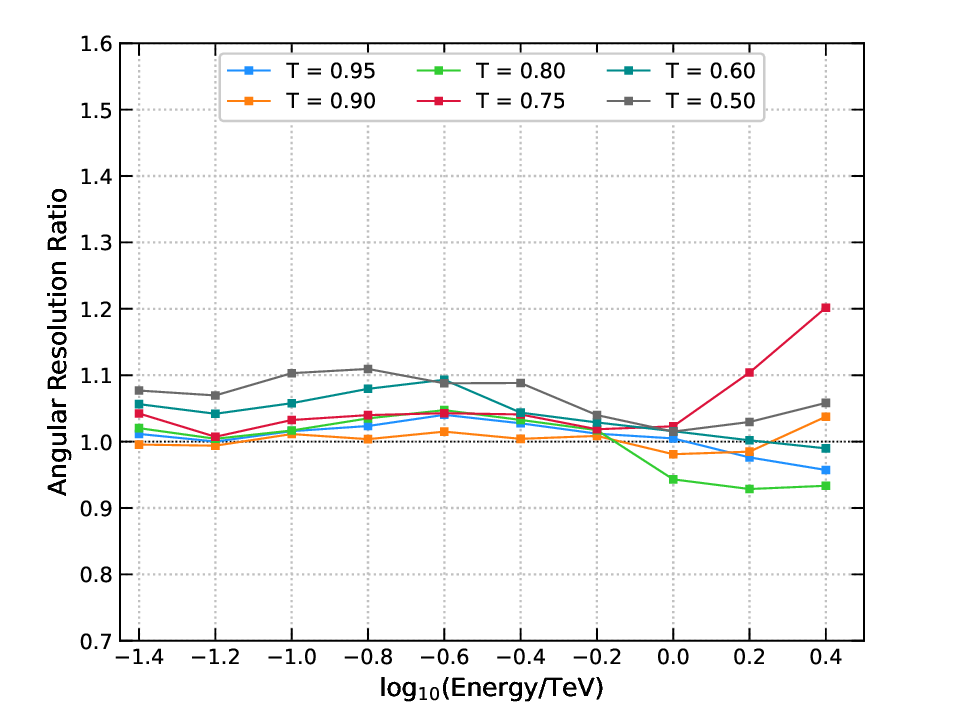}
            }
            \subfigure[13 km a.g.l.] 
            {
                \includegraphics[width=.45\textwidth]{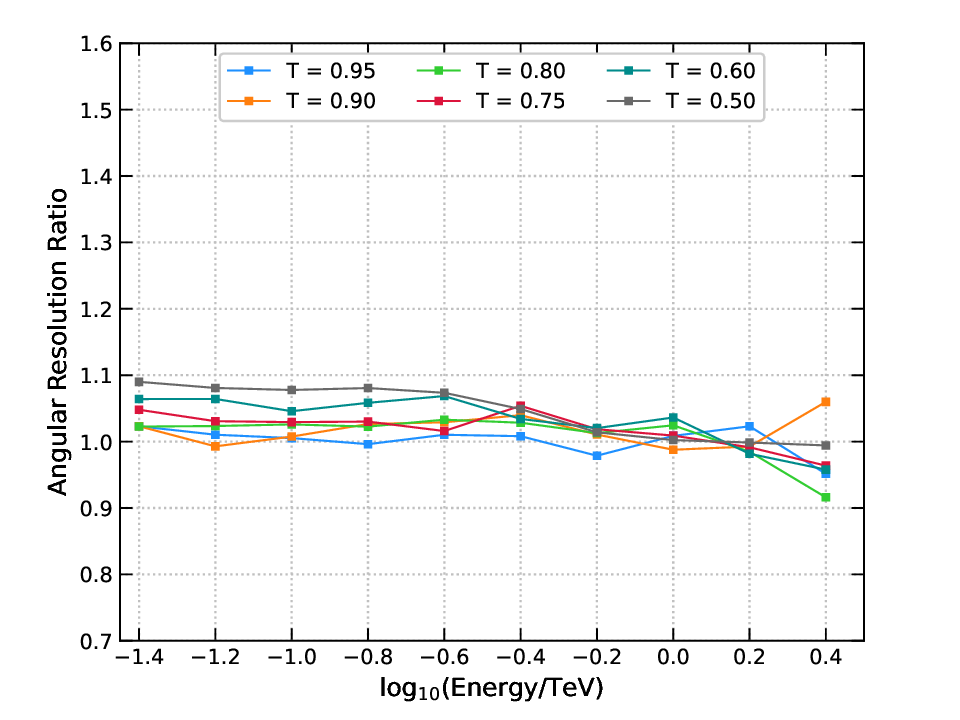} 
            }
            
        \caption{Angular resolution ratio as a function of energy for the subarray of 4 LSTs - a comparison between the cloudless atmosphere and the atmosphere with clouds. The lower value corresponds to better resolution.}
        \label{fig:subfigsLSTsAnRes}
        \end{figure*}

The impact of clouds on the angular resolution of 4 LSTs (figure \ref{fig:subfigsLSTsAnRes}) is rather small over the whole energy range, as the angular resolution is worse by only 10\% or less compared to the case where no clouds are present. For example, in the presence of clouds at 3 km a.g.l., the angular resolution varies from 0.18-0.21 degrees at an energy of $\approx 40$ GeV to 0.05 degrees at an energy of 2.5 TeV, depending  on the transmission.

\begin{figure*}[t]
        \centering
            \subfigure[3 km a.g.l.]
            {
                \includegraphics[width=.45\textwidth]{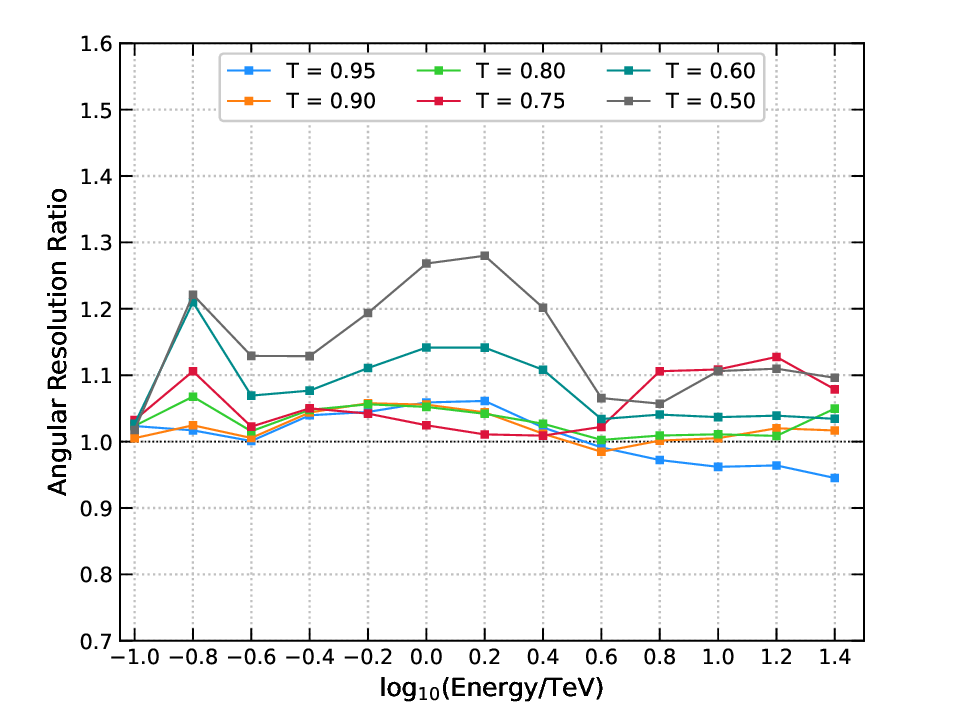}
            } 
            \subfigure[5 km a.g.l.] 
            {
                \includegraphics[width=.45\textwidth]{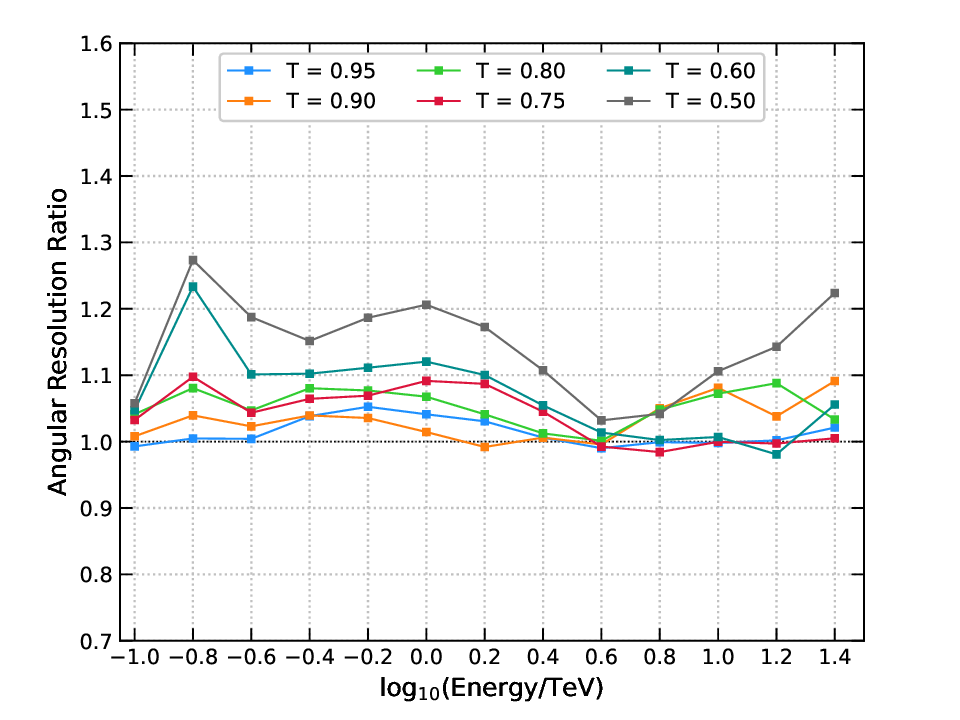}
            }\\
            \subfigure[7 km a.g.l.] 
            {
                \includegraphics[width=.45\textwidth]{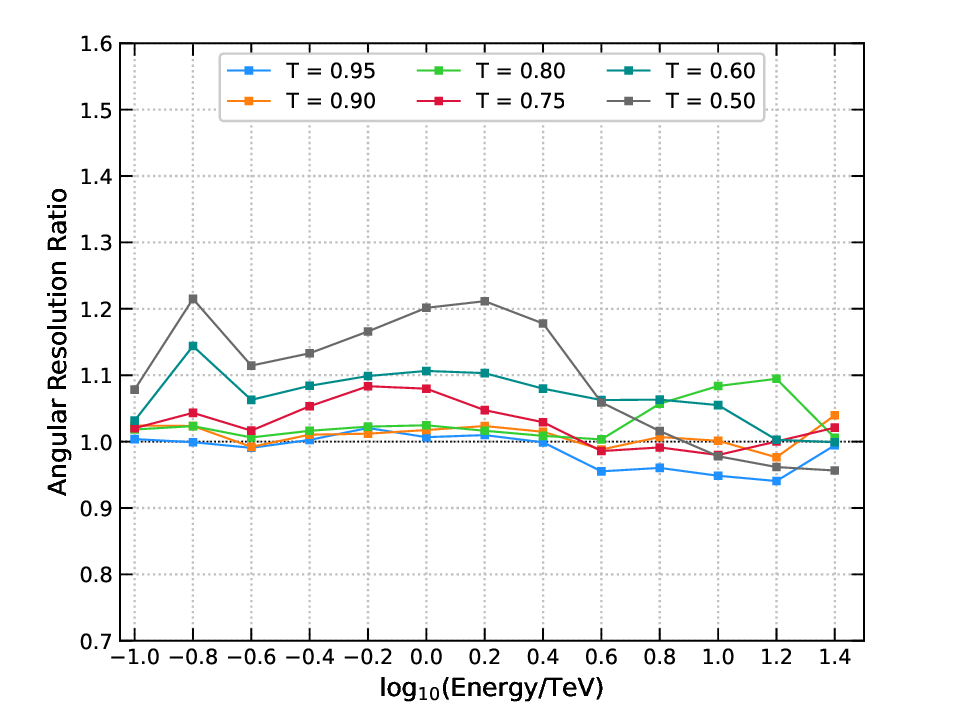} 
            } 
            \subfigure[9 km a.g.l.] 
            {
                \includegraphics[width=.45\textwidth]{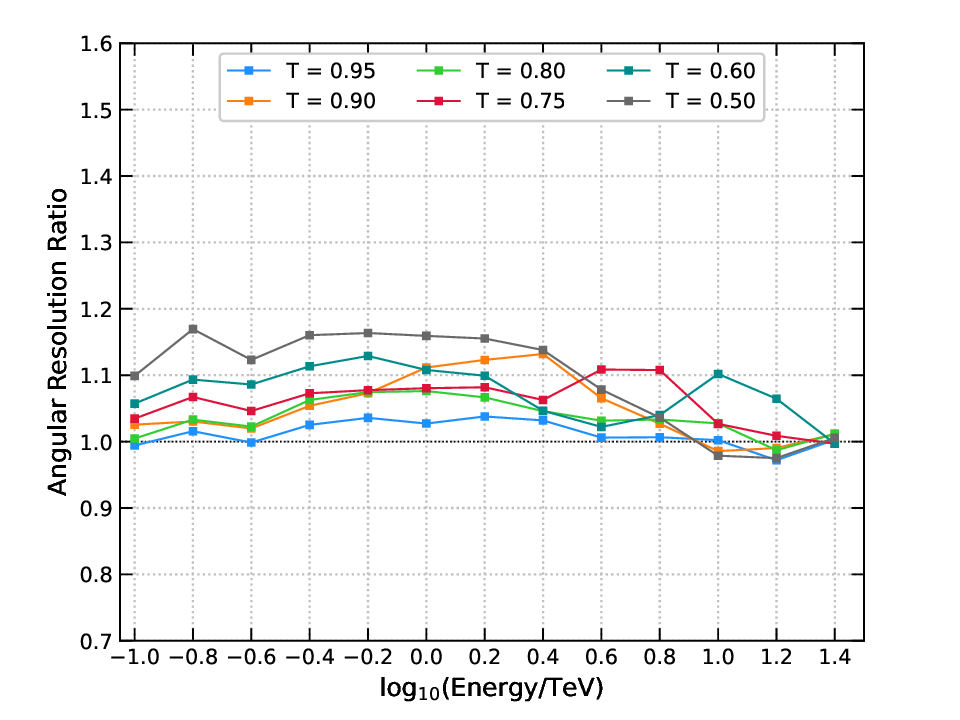} 
            }\\
            \subfigure[11 km a.g.l.] 
            {
                \includegraphics[width=.45\textwidth]{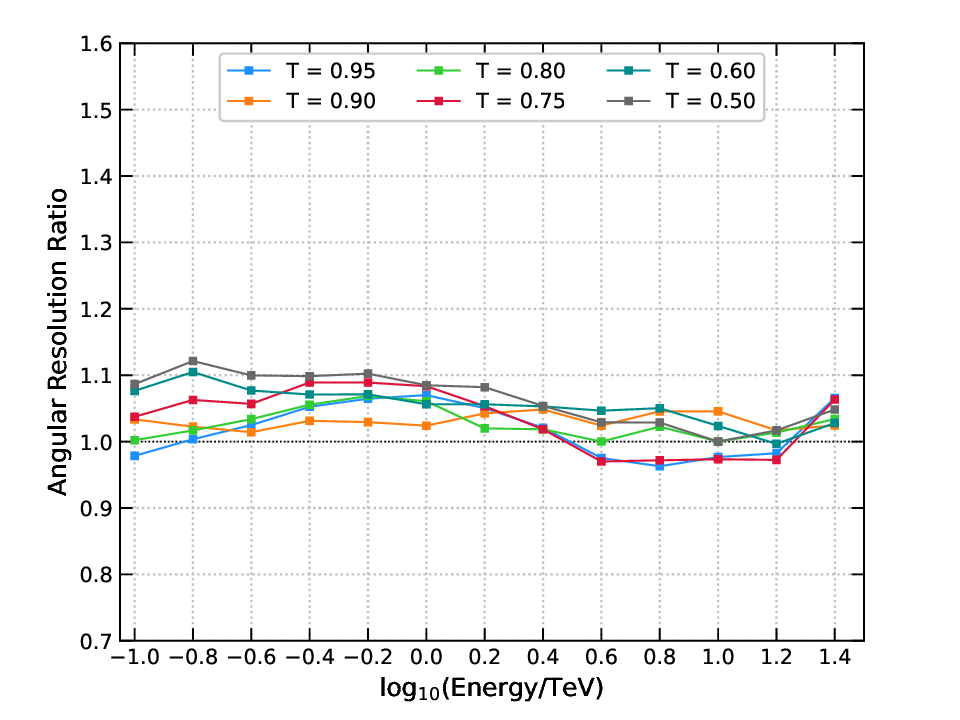} 
            }
            \subfigure[13 km a.g.l.] 
            {
                \includegraphics[width=.45\textwidth]{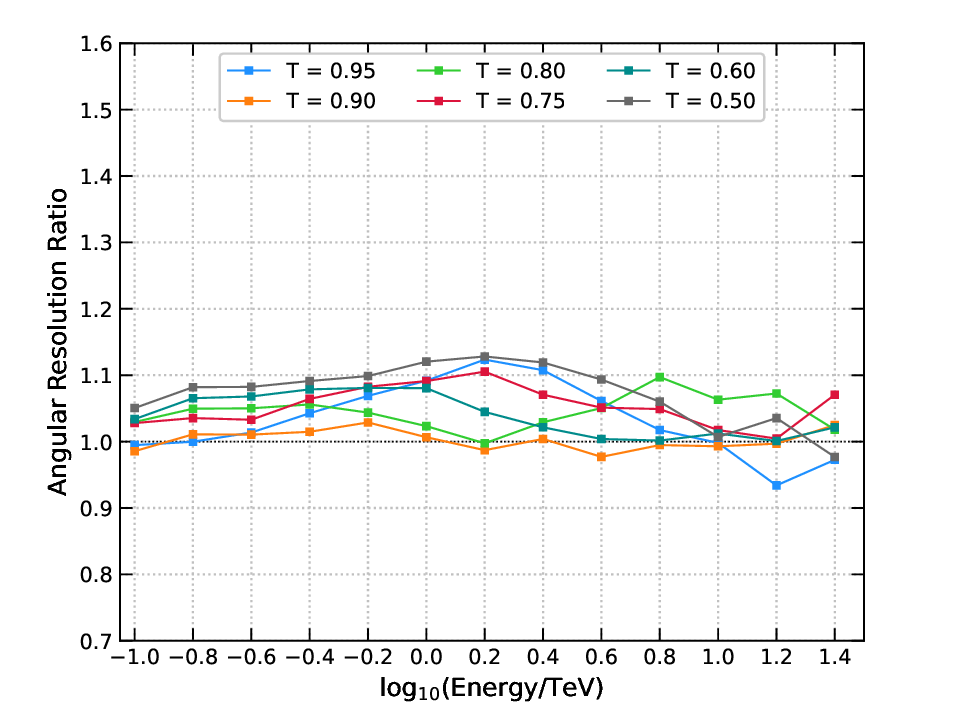}
            }
        \caption{Angular resolution ratio as a function of energy for the subarray of 15 MSTs - a comparison between the cloudless atmosphere and the atmosphere with clouds. The lower value corresponds to better resolution.}
        \label{fig:subfigsMSTsAnRes}
        \end{figure*}

In the case of 15 MSTs, the angular resolution improves with energy, reaching a plateau above 2.5 TeV (figure \ref{fig:subfigsMSTsAnRes}), and in general the cloud contribution is negligible at these energies ($\lesssim$ 10\%). At lower energies, the angular resolution is worsened by up to 30\% ($T\leq 0.75$) compared to the clear atmosphere, although the effects become weaker with increasing altitude and transmission. For more transparent clouds ($T\geq 0.90$), the angular resolution is $\leq 5\%$ worse. For clouds at 3 km altitude, the angular resolution varies with transmission from $\approx$ 0.14 deg to 0.11 deg at an energy of $\approx 150$ GeV to 0.04 deg at an energy of 25 TeV.

\begin{figure*}[t]
        \centering
            \subfigure[3 km a.g.l.]
            {
                \includegraphics[width=.45\textwidth]{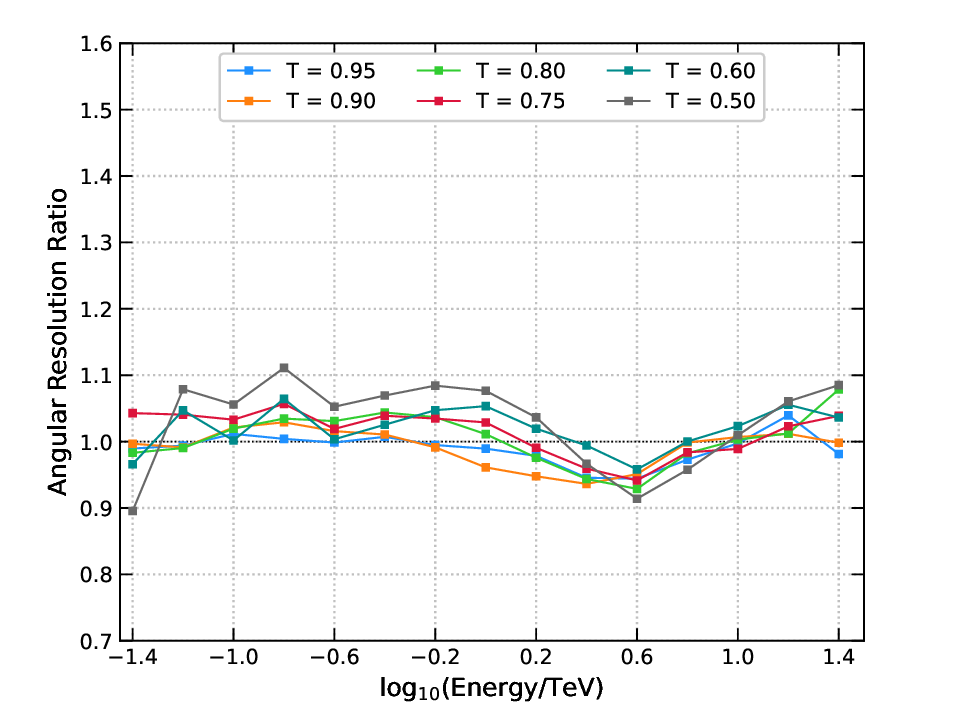}
            } 
            \subfigure[5 km a.g.l.] 
            {
                \includegraphics[width=.45\textwidth]{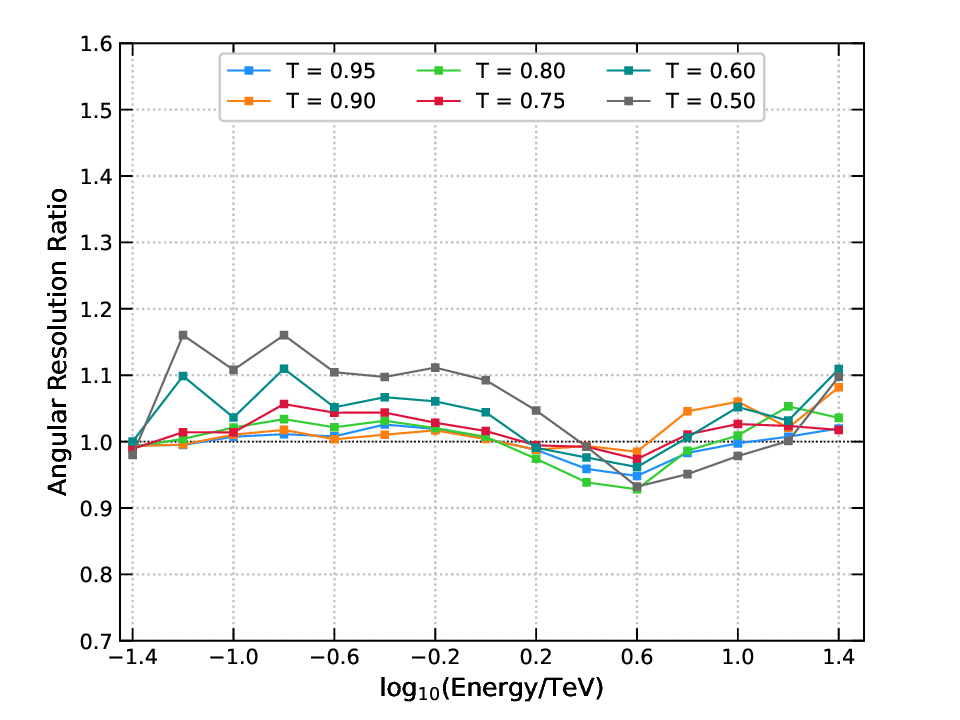}
            }\\
            \subfigure[7 km a.g.l.] 
            {
                \includegraphics[width=.45\textwidth]{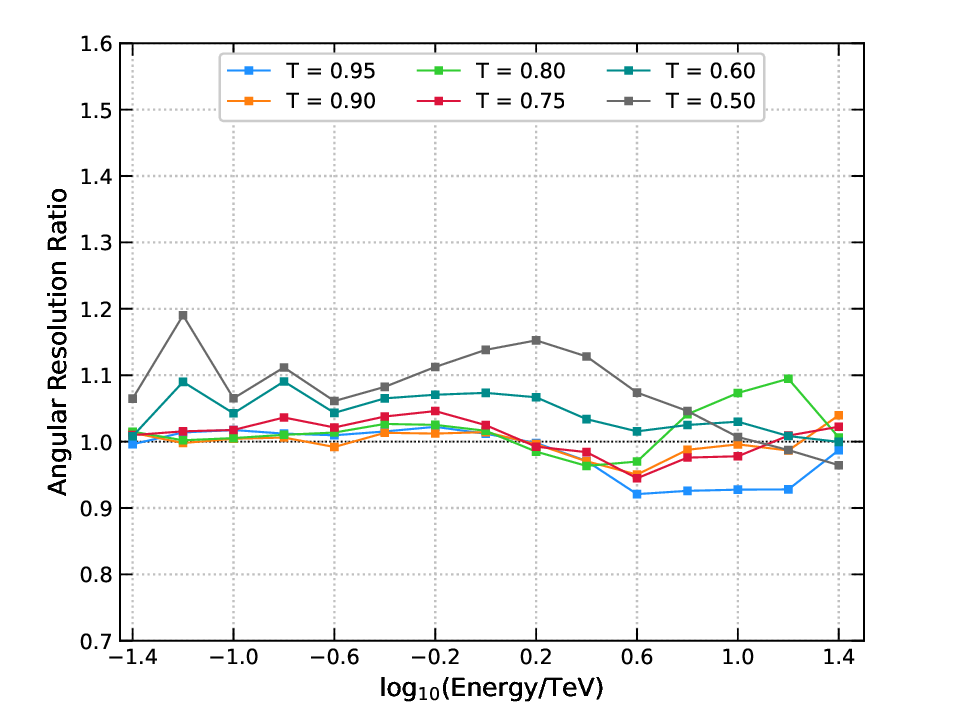} 
            } 
            \subfigure[9 km a.g.l.] 
            {
                \includegraphics[width=.45\textwidth]{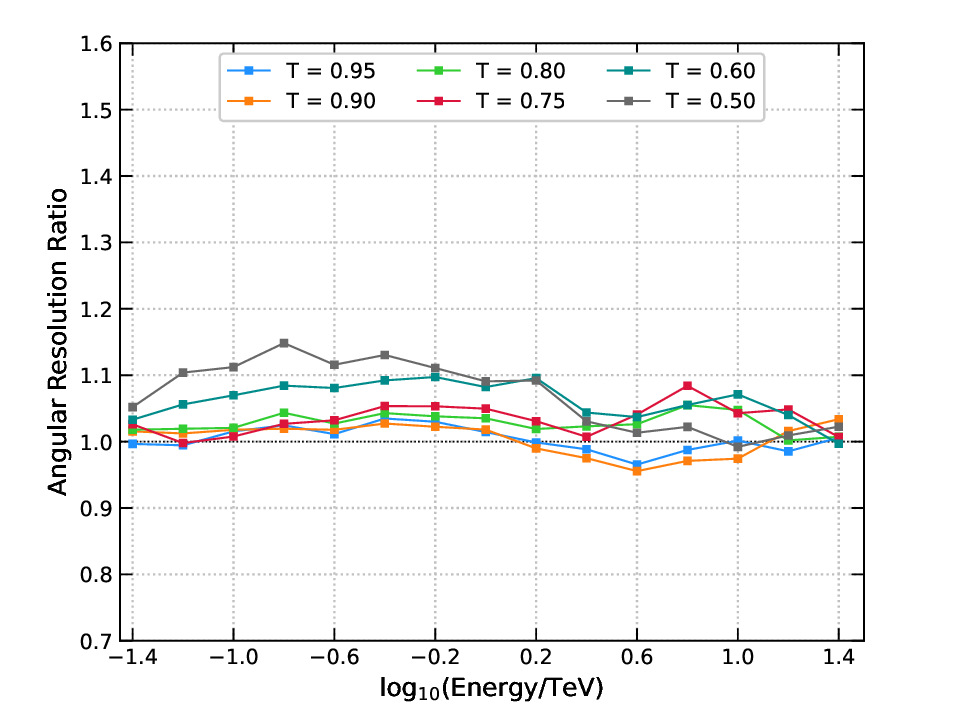} 
            }\\
            \subfigure[11 km a.g.l.] 
            {
                \includegraphics[width=.45\textwidth]{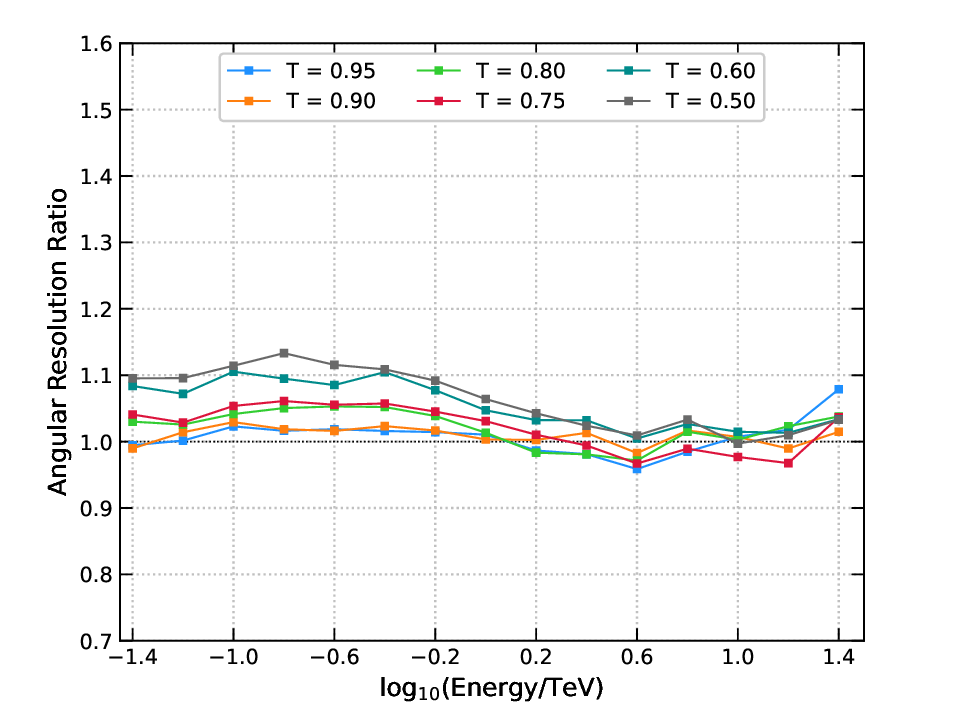} 
            }
            \subfigure[13 km a.g.l.] 
            {
                \includegraphics[width=.45\textwidth]{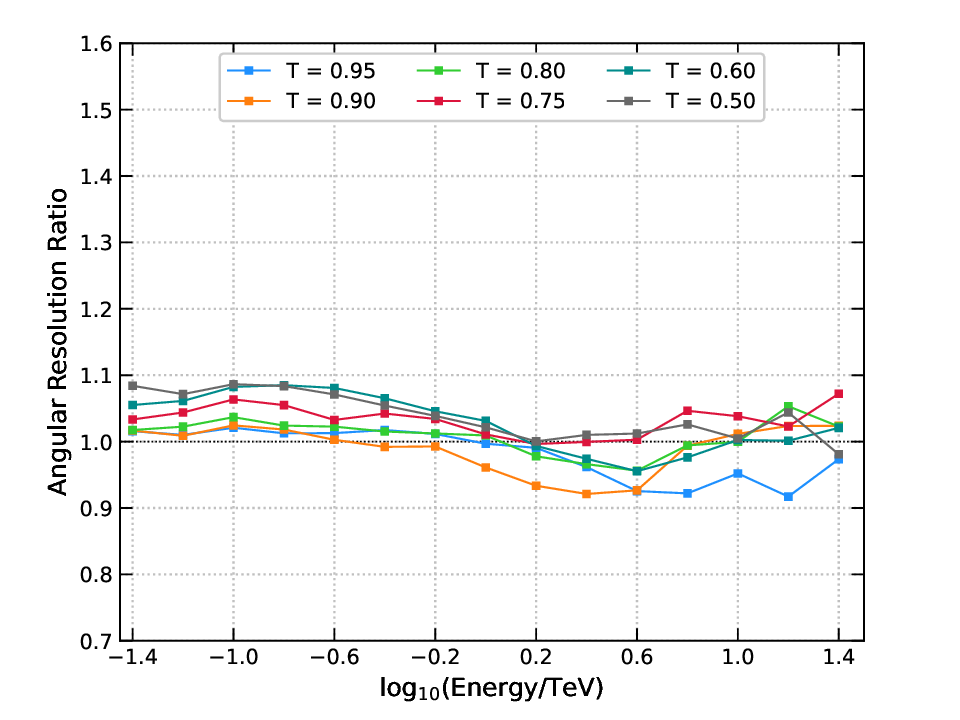}
            }
        \caption{Angular resolution ratio as a function of energy for the CTA-N array - a comparison between the cloudless atmosphere and the atmosphere with clouds. The lower value corresponds to better resolution.}
        \label{fig:subfigsCTAAnRes}
        \end{figure*}

Although reconstruction of direction is improved for the CTA-N (figure \ref{fig:subfigsCTAAnRes})  compared to individual subarrays, the angular resolution is still degraded for clouds of transmission below 0.60 at energies below 1 TeV.

\newpage
\clearpage

\section{Summary and conclusions}
A large-scale database of simulated extensive air showers induced by gamma rays, protons, and electrons was generated using CORSIKA code, as well as the emitted Cherenkov light in the wavelength range from 240 nm to 700 nm. Simulated showers were subjected to further simulations of the telescope response using \texttt{sim\_telarray} software to assess the Cherenkov Telescope Array response in the presence of clouds based on customized atmospheric transmission models produced with MODTRAN software. A total of 36 transmission models were produced, differing in transmission (optical depth) and cloud height.

The presence of clouds may severely degrade the performance of telescopes depending on the cloud properties. The most significant influence on the performance of CTA in the conditions of reduced atmospheric transmission, and its respective subarrays,  arises from the density of the cloud, i.e., its optical depth. The increased optical depth of the medium decreases the effective area of telescopes and, subsequently, increases the energy threshold. In the presence of higher clouds ($\geq$ 11 km), the influence is small, even for the most opaque clouds. The latter is well understood by the fact the bulk of air showers have a maximum below the clouds' bases, and only a small portion of photons is lost. The largest differences are evident near the energy threshold, while the effects become more stable at higher energies. Above the energy threshold of both types of telescopes ($\geq$ 150 GeV), clouds have a similar impact on individual subarrays and full Northern Cherenkov Telescope Array performance. Compared to the performance in the clear atmosphere, lower clouds worsen the sensitivity by up to 60\% at the energy threshold and by up to 20\% above 1 TeV.

Energy resolution in the presence of lower clouds ($\leq$ 7 km a.g.l.) is worse by 30\% at the energy threshold and by less than 10\% at energies above 1 TeV. In the energy range common to both types of telescopes (between 100 GeV and 1 TeV), the performance degradation of the energy estimator is more severe for 15 MSTs, as these telescopes are less sensitive to lower energies. However, the performance of both subarrays is similar for higher clouds.

The reconstruction of direction is well managed even in the presence of low and very opaque clouds, as the deviations from the shower axis are small. Angular resolution is worse by 20\% and 30\% for subarray of, respectively, 4 LSTs and 15 MSTs. Even though the angular resolution is improved for CTA-N compared to individual subarrays, the relative difference in the angular resolution is still around 10\%. 

To summarize the influence of different optical depths and altitudes on the sensitivity of Cherenkov telescopes, a simple semi-analytical model of sensitivity degradation has been introduced.  Provided model is consistent with Monte Carlo simulations within an accuracy of $\lesssim10\%$, except in the case of very low clouds (3 km a.g.l.), where the model is not able to reproduce the slight gain of the performance with decreasing height of the cloud. 

\acknowledgments
This work has been funded by the University of Rijeka grant \texttt{uniri-prirod-18-48}. Simulations have been performed using the supercomputer Bura within the Center for Advanced Computing and Modelling, University of Rijeka.
Julian Sitarek is supported by Narodowe Centrum Nauki grant number  \texttt{2019/34/E/ST9/00224}. The authors would like to thank Gernot Maier from the DESY Research Center, Germany, for the help with the MODTRAN software.

\newpage
\bibliographystyle{JCAP}
\bibliography{references}

\end{document}